\documentclass[a4paper,fleqn,usenatbib]{mnras}

\usepackage[T1]{fontenc}
\usepackage{ae,aecompl}
\usepackage{scrextend}
\usepackage{multirow}
\usepackage{pifont}
\usepackage[normalem]{ulem}
\usepackage{graphicx}	
\usepackage{amsmath}	
\usepackage{amssymb}	
\usepackage{xcolor}

\title[\textit{u}-band calibration]{Photometric calibration in \textit{u}-band using blue halo stars}

\author[S. Liang et al.]{
Shuang Liang,$^{1,2}$\thanks{E-mail: sliang92@stanford.edu}
Anja von der Linden$^{1}$
\\
$^{1}$Department of Physics and Astronomy, Stony Brook University, {\it Stony Brook, NY 11790}\\
$^{2}$Kavli Institute for Particle Astrophysics and Cosmology, Department of Physics, Stanford University, {\it Stanford, CA 94305}
}

\date{Accepted XXX. Received YYY; in original form ZZZ}

\pubyear{2022}

\usepackage{newtxtext,newtxmath}

\begin{document}
\label{firstpage}
\pagerange{\pageref{firstpage}--\pageref{lastpage}}
\maketitle

\begin{abstract}
We develop a method to calibrate \textit{u}-band photometry based on the observed color of blue galactic halo stars. 
The galactic halo stars belong to an old stellar population of the Milky Way and have relatively low metallicity. The ``blue tip'' of the halo population --- the main sequence turn-off (MSTO) stars --- is known to have a relatively uniform intrinsic edge \textit{u-g} color with only slow spatial variation. In SDSS data, the observed variation is correlated with galactic latitude, which we attribute to contamination by higher-metallicity disk stars and fit with an empirical curve. This curve can then be used to calibrate \textit{u}-band imaging if \textit{g}-band imaging of matching depth is available. Our approach can be applied to single-field observations at $|b| > 30^\circ$, and removes the need for standard star observations or overlap with calibrated \textit{u}-band imaging. We include in our method the calibration of \textit{g}-band data with ATLAS-Refcat2. 
We test our approach on stars in KiDS DR 4, ATLAS DR 4, and DECam imaging from the NOIRLab Source Catalog (NSC DR2), and compare our calibration with SDSS. For this process, we use synthetic magnitudes to derive the color equations between these datasets, in order to improve zero-point accuracy.
We find an improvement for all datasets, reaching a zero-point precision of 0.016 mag for KiDS (compared to the original 0.033 mag), 0.020 mag for ATLAS (originally 0.027 mag), and 0.016 mag for DECam (originally 0.041 mag). 
Thus, this method alone reaches the goal of 0.02 mag photometric precision in \textit{u}-band for the Rubin Observatory's Legacy Survey of Space and Time (LSST).
\end{abstract}




\section{Introduction}
The near-ultraviolet wavelength regime accessible from the ground - in current facilities typically the \textit{u}-band\footnote{In this introduction, we will use the term \textit{u}-band generically; our analysis uses specific \textit{u}-band filters, but can be easily adapted to other filters limited by the atmosphere in the NUV.} - carries important astrophysical information, but is notoriously difficult to calibrate due to the steep cut-off in atmospheric transparency.
For example, \textit{u}-band observations are important to
the study of stellar metallicity and stellar formation/evolution \citep{u_metal,u_evolv}, variable stars \citep{u_var1, u_var2}, galactic archaeology \citep{GA}, and composition of asteroids \citep{kbo}. The \textit{u}-band measurement
also benefits photometric redshift estimates by being able to break low- vs. high-redshift degeneracies \citep{ly1996a}, thus is important for cosmological imaging surveys such as the upcoming Legacy Survey of
Space and Time (LSST) to be conducted at the Vera C. Rubin Observatory.
With great care, \textit{u}-band photometric zero-points of large contiguous sky areas can be calibrated to roughly half the precision of optical bands; the Sloan Digital Sky Survey \citep[SDSS;][]{sdss, dr15} reached 13 mmag in \textit{u}-band compared to 8, 8, and 7 mmag in the \textit{gri}-bands \citep{ubercal, hypercal}, and has the potential to be calibrated twice as good \citep{stripe82}.  The minimum requirement for Rubin-LSST is 15 mmag in the \textit{grizy}-bands (the design spec is 10 mmag, and the stretch goal is 5 mmag), while the \textit{u}-band may be worse but with less than a factor of 2 (table~15 of LPM-17\footnote{\url{https://docushare.lsst.org/docushare/dsweb/Get/LPM-17}}). 

\textit{u}-band calibration of pointed single-field observations with no overlapping calibrated survey data is even more difficult.  While in the northern extragalactic sky, \textit{u}-band images can be calibrated by comparing with SDSS, the current southern sky surveys lack in depth and precision. SkyMapper\footnote{\url{http://skymapper.anu.edu.au/}} covers much of the Southern sky but is comparatively shallow.  Moreover, its {\it u}-band is notably different from the other {\it u}-bands discussed here, resulting in large color terms relative to other filter sets, and accordingly large zero-point scatter when comparing to other observations \citep{skymapper}.
The ATLAS survey \citep[][]{atlas} reaches 22.0 magnitude at 5$\sigma$ in the \textit{u}-band with a zeropoint precision of 0.035 mag, but only covers about 4000 deg$^2$ (DR4).  The Kilo-Degree Survey\footnote{\url{http://kids.strw.leidenuniv.nl/}} \citep[KiDS;][]{kids} has a 5$\sigma$ limiting magnitude as deep as 24.2 in the \textit{u}-band with 0.035 mag zero-point precision but covers only $\sim 1000$ deg$^2$ as of DR4.

When no calibrated reference stars are available in the field of view (or those stars saturate), the \textit{u}-band calibration becomes very challenging. Typically, exposures
of other fields containing calibrated standard stars are required for the calibration, with the assumption that the atmospheric conditions do not vary between the fields of interest
and the standard star fields. As the \textit{u}-band transmission function is largely defined by the atmosphere, this assumption becomes questionable. As a result, the traditional standard star calibration method introduces considerable intrinsic uncertainties.

One calibration approach that circumvents the need for extrapolating from standard star fields is to use the observed colors of the stellar locus \citep{slr_2, slr_3, big-macs, desy3_slr, s_color, s_color_cali}, namely that the colors of stars fall on a well-defined line in color-color diagrams.
It is a very powerful calibration tool for broad-band photometry of extragalactic fields, and is especially useful for single-field observations in optical bands, where it can deliver zero-points of $\le 0.02$ mag precision \citep{big-macs}. However, it is less effective for \textit{u}-band, since variation in stellar metallicity leads to a widening of the stellar locus due to metal absorption lines in the NUV (see App.~\ref{sec:appe}). The imprint of metallicity on the \textit{u-g} color is typically 10 times larger than on the \textit{g-r} color \citep{slr_3}. It is generally difficult to correct for the \textit{u}-band scatter in stellar locus because the metallicity distribution of observed stars is usually unknown. 

We here develop an alternative way to calibrate \textit{u}-band photometry of single-field observations without overlapping calibrated \textit{u}-band catalogs, by isolating a population of stars with homogeneous metallicity.  Namely, our method utilizes the observed \textit{u-g} color of halo main sequence turn off (MSTO) stars, which are F/G type stars in the galactic halo that typically have a relatively low metallicity. Based on SDSS data, \citet{blue-tip} find that the intrinsic \textit{u-g} color of the very edge of the distribution --- those ``blue tip'' stars --- is roughly constant over the SDSS footprint (mostly the Northern sky) with only small spatial variation. We observe that the remaining spatial variation is strongly correlated with galactic latitude $b$, which can be well descried by a plane-parallel model. We fit this model across on the entire SDSS footprint, and use it as the basis of a method of \textit{u}-band zero-point calibration for any instrument. Assuming that
the spatial relation is true in the Southern sky as well, it can be used to calibrate data sets beyond the SDSS coverage. The method requires the exposures to be deep enough to reach the halo MSTO stars
($g\approx 21$), and \textit{g}-band imaging to be available as well.  The latter do not need to be previously calibrated, as we provide a prescription to do so using the ATLAS-Refcat2 catalogue \citep{refcat}.

As a proof of principle, we here test our calibration method on KiDS, ATLAS and the NOIRLab Source Catalog \citep[NSC,][]{nsc} stars, and
compare the results with the SDSS \textit{u}-band photometry. We find an improvement in the zero-point scatter in all cases: the zero-point precision of \textit{u}-band improves from 0.033 mag to 0.016 mag for KiDS, 0.027 mag to 0.020 mag for ATLAS, and 0.041 mag to 0.016 mag for NSC. The quoted after-calibration scatters are all based on a joint \textit{ug}-bands calibration, as our method calibrates the \textit{u}-band relative to the \textit{g}-band. We show that a joint calibration is necessary, by examining the original \textit{g}-band photometry of KiDS, ATLAS and NSC catalog and estimating their contribution to the uncertainties in the final calibrated \textit{u}-band. We show that the \textit{g}-band photometry can be improved by calibrating with respect to ATLAS-Refcat2 \citep[The Asteroid Terrestrial-impact Last Alert System]{refcat}, and compare the calibration with SDSS and PanSTARRS \citep[The Panoramic Survey Telescope and Rapid Response System;][]{pan}. In sum, the method presented here can be used to calibrate {\it ug}-band imaging, using only overlapping ATLAS-Refcat2 photometry. It is thus available for the full sky; however, since it is based on colors of halo stars, we limit the analysis to galactic latitudes $|b|>30^\circ$ and with extinction {\it E(B-V)}$<0.1$. 

This paper is organized as follows. In Sect.~\ref{sec:data} we describe the data sets used in this work, together with the selections of the star samples and the details of catalog cross-matching. Sect.~\ref{sec:model} details our method of detecting the ``blue tip'' \textit{u-g} color and the fitting of the color-latitude model. In Sect.~\ref{sec:rec} we present the re-calibration on KiDS, ATLAS and NSC. We discuss possible applications and summarize our work in Sect.~\ref{sec:dis}. App.~\ref{sec:appa} details the data acquirement and selection. The color equations among surveys are derived in App.~\ref{sec:appb} on the original photometry, and in App.~\ref{sec:appc} on synthetic magnitudes. In App.~\ref{sec:appd} we attempt to improve the calibration by rejecting disk stars with Gaia EDR3. In App.~\ref{sec:appe} we compare our approach with stellar locus based methods. Quoted magnitudes are in the AB system \citep{ab_mag}.

\section{Data Description}
\label{sec:data}

\begin{figure*}
	\centering
        \includegraphics[width=0.99\hsize]{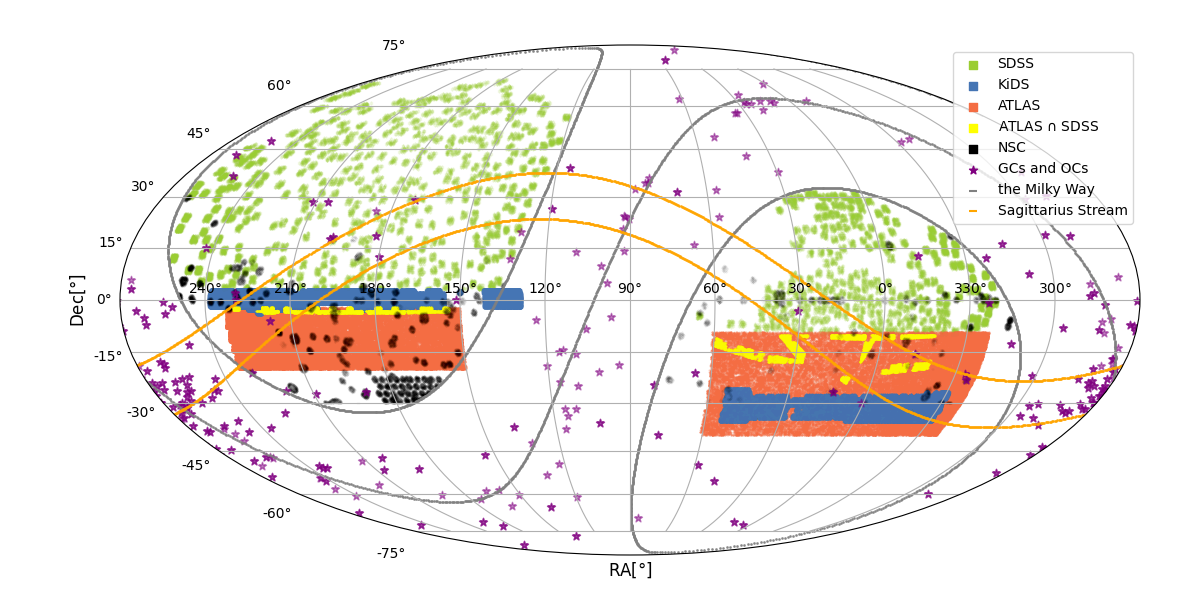}
	\caption{Sky coverage of the data used in this work. For each survey, we indicate the fields that we use ($1\times1$~deg$^2$ or $2\times2$~deg$^2$) by colored boxes: green for SDSS, blue for KiDS, red for ATLAS, and gray for NSC. SDSS fields are selected randomly from the area with $|b|>30^\circ$ and E(B-V)$<0.1$. The KiDS North area entirely overlaps SDSS, while KiDS South has no SDSS overlap.
	ATLAS tiles overlapping SDSS are shown in yellow. PanSTARRS is available for all fields at declination above $-30^{\circ}$, and ATLAS-Refcat2 is full-sky. The star symbols indicate some known Globular Clusters (GCs) and Open Clusters (OCs), which we exclude here. The Milky Way is indicated by gray lines at  $b=30^{\circ},0^{\circ},-30^\circ$. The Sagittarius Stream is indicated as orange lines at $|\beta| = 7^\circ$, where $\beta$ is the latitude in the Sagittarius Coordinate System \citep{Sag}.} 
	\label{fig:cover}
\end{figure*}

In this work, we will test the calibration methods (for both the \textit{u}- and \textit{g}-bands) using star catalogs from KiDS \citep{kids}, ATLAS \citep{atlas} and NSC \citep{nsc}. They will be referred to as the validation sets. We use much of the large area of SDSS \citep{sdss} to develop the calibration approach for \textit{u}-band, and use other parts of SDSS (regions overlapping the validation sets) as the truth for validation. ATLAS-Refcat2 \citep{refcat} is used for calibrating the \textit{g}-band. Since the latest SDSS calibration utilizes PanSTARRS calibration \citep{hypercal},  we also compare our \textit{g}-band calibration to PanSTARRS directly. The sky coverage of the data sets used in this work is shown in Fig.~\ref{fig:cover}. Table~\ref{tab:purpose} summarises the usage of these data sets.  
\begin{table}
	\caption{Overview of the datasets used in this work. We develop our \textit{u}-band calibration model with SDSS and apply the method to the validation sets. SDSS is also used as the \textit{ug}-bands truth catalog for the validation sets. Notice that the two SDSS catalogs are selected to not overlap each other. We also develop a \textit{g}-band calibration procedure to accompany the \textit{u}-band calibration, based on ATLAS-Refcat2 and using PanSTARRS as the truth for validation.}
	\label{tab:purpose}
	\centering
	\begin{tabular}{|r|c|}
		\hline
		Model Construction     &  SDSS(\textit{u}), ATLAS-Refcat2(\textit{g}) \\
		Validation Sets        &  KiDS, ATLAS, NSC \\
		Truth Catalogs         &  SDSS(\textit{ug}), PanSTARRS(\textit{g}) \\
        \hline
    \end{tabular}
\end{table}

For developing the method, we strive to select stellar samples of high purity. Therefore, we use the photometric quality flags provided by SDSS in all available bands, i.e. {\it ugriz}. When calibrating {\it u}-band data with this method, similar care should be taken to select stellar samples with high purity, even if fewer bands are available. We expect that the single-field datasets that this method is of interest to are typically deeper than the survey data examined here, such that a robust selection of stars to this depth ($g\approx 21$) should be straightforward. 

We initially keep only stars with magnitude 14-21 in all available bands and uncertainties $<0.1$, largely to illustrate how we develop this method. We later place an explicit cut of $18.5<g<20$ which obviates the previous cuts. We use extinction-corrected magnitudes; where those are not explicitly given in a catalog, we calculate them from \citet{sfd}. Details of the sample selection for each dataset can be found in App.~\ref{sec:appa}.

\subsection{SDSS data}
We use stars from SDSS DR 15 \citep{dr15}. We randomly select 640 non-overlapping fields of $2\times2$ deg$^2$:
in the galactic North ($b>30^\circ$), we query 400 fields from the region 8h$<$R.A.$<$16h and $-10^\circ<$ Dec. $<70^\circ$; in the South ($b<-30^\circ$), we select 240 fields in between $-3$h$<$R.A.$<$5h and $-10^\circ<$ Dec. $<30^\circ$. We exclude the KiDS and ATLAS regions, as well as tiles that overlap known Globular and Open Clusters. 

\subsection{KiDS data}
KiDS is a 1350 deg$^2$ optical imaging survey designed to study weak gravitational lensing and galaxy evolution.
We use the KiDS Data Release 4.1 \citep{kids} multi-band source catalogue. The original KiDS photometry is calibrated based on stellar locus colors, followed by a latitude-dependent correction for \textit{u}-band (similar to our approach), and a final calibration of \textit{r}-band magnitudes to Gaia DR2 \citep{gaia_mission}, shifting the zero-points of all bands accordingly \citep{kids}. The catalogue consists of sources detected from 1006 tiles ($\sim 1\times 1$ deg$^2$ FOV) observed with filters \textit{u,g,r,} and \textit{i}, all of which have low extinction {\it E(B-V)}$<0.1$. The entire North part of KiDS (492 tiles) overlaps SDSS, while the South part (514 tiles) is completely out of the SDSS footprint. Notice that a small portion of KiDS North has a galactic latitude lower than $30^\circ$ (see Fig.~\ref{fig:cover}), but since the extinction in that region is low, we do not exclude those tiles. We use the extinction-corrected Gaussian aperture and PSF (GAaP) magnitudes, a measurement of the flux inside a tapered aperture. The GAaP magnitudes account for PSF differences between observations in different filter bands for accurate color measurements while optimizing SNR. For stars, the GAaP magnitudes measure the total flux \citep{gaap}.  

\subsection{ATLAS data}
\label{subsec:atlas_data}
ATLAS is an optical \textit{ugriz} survey that aims to cover 4700 square degrees of the Southern Sky at high galactic latitudes to comparable depths as SDSS in the North. ATLAS utilizes the same telescope as KiDS (VST, the VLT Survey Telescope) thus they share the same set of filters. The \textit{g}-band photometry is calibrated with Gaia DR2 \cite{gaia_dr2}, and the \textit{u}-band is calibrated with overlap photometry \citep{atlas_web}.
We download sources in ATLAS Data Release 4\footnote{\url{http://osa.roe.ac.uk/}} from two regions: the galactic North region $10$h$<$R.A.$<$16h and $-4^\circ<$ Dec. $<0^\circ$, and the galactic South with $-2.5$h$<$R.A.$<$4h. We split the sources into 1019 tiles with an approximate tile size of $2\times 2$ deg$^2$, a larger size than KiDS in order to balance the number of stars in each tile. Most of the ATLAS regions overlapping SDSS are at the edge of SDSS, yielding fewer stars matching with SDSS. Out of all the 1019 tiles, only 152 tiles overlap a large enough SDSS area to yield $>300$ stars in common). Notice that in the Galactic North region, we limit the data to Dec. $>-4^\circ$ in order to remain in the fully contiguous area of SDSS coverage, avoiding single scans at high airmass \citep[see e.g. fig.~1 of ][]{hypercal}. 

\subsection{NSC data}
\label{subsec:nsc_data}

We use the DECam {\it ug} data from the NOIRLab Source Catalog (NSC), a collection and reduction of public imaging data taken from various facilities \citep{nsc}. Only a few NSC exposures are taken with the \textit{u}-band filter, all taken with DECam.
The NSC DR2 catalogue calibrates the \textit{u}-band imaging with SkyMapper and Gaia DR2 in the South, and GALEX, Gaia and 2MASS \citep{2mass} in the North.
From the NSC catalog, we select stars at lower Galactic Latitude ($|b|>30^\circ$) and overlapping PanSTARRS
(Dec. $>-30^\circ$), and assign them to individual DECam exposures according to exposure centers. 
Only 158 tiles have enough stars (more than 300 within $18.5<g<20$ and \textit{ug} mag error $< 0.1$) 
and only 72 of them have enough matches with SDSS. 

\subsection{PanSTARRS data}
PanSTARRS is a wide-field survey with five broadbands ({\it grizy}). PanSTARRS has a consistent photometric calibration of $7 \sim 12.4$ mmags across all of the Northern sky down to Dec. $>-30^\circ$ \citep{pan_cali}.
We retrieve PanSTARRS data to cover the footprint of the validation sets.

\subsection{ATLAS-RefCat2 data}
\label{sec:refcat data}
The ATLAS-Refcat2 (in the following Refcat2) is an all-sky star catalog containing nearly one billion stars which is 99\% complete
down to magnitude $< 19$ \citep{refcat}. The catalog is a compilation of PanSTARRS DR1, SkyMapper DR1 \citep{skymapper}, Tycho-2 \citep{tycho}, the Yale Bright Star Catalog \citep{BSC}, new detection from ATLAS Pathfinder and re-processed APASS images \citep{apass}, in
order to provide a consistent reference photometry catalog for the Asteroid Terrestrial-impact Last Alert System \citep[ATLAS,][]{other_atlas}. Gaia DR2 \citep{gaia_dr2} and 2MASS \citep{2mass} are used for zero-point homogeneity over the entire sky. Source photometry is converted into the PanSTARRS {\it griz} system and then combined with a weighted-mean to give the final Refcat2 magnitudes. The overall RMS uncertainty of Refcat2 is $<5$ mmags, where in some small patches the uncertainty is $<20$ mmags. We correct for the extinction with the same band coefficients as PanSTARRS. 

We use Refcat2 as the {\it g}-band reference catalog instead of Gaia, because the Refcat2 {\it g}-band filter is closer to our target data sets. The Gaia \{$G, G_{\rm BP}, G_{\rm RP}$\} filter system yields a more complicated color equation.

\subsection{Catalog Cross-Matching}
\label{subsec:match}

We match individual stars across the catalogs for two purposes: 
\begin{enumerate}
    \item to calibrate \textit{g}-band magnitudes with respect to Refcat2, and
    \item to compare the calibrated magnitudes (\textit{ug}-bands) to the truth datasets (SDSS/PanSTARRS).
\end{enumerate}
For both purposes, we need to match the validation sets to other surveys, i.e. to Refcat2 as the {\it g}-band calibrator and to SDSS/PanSTARRS as the truth.  The match radius is set to 1 arcsec. Table~\ref{tab:star_num} shows the average number of stars per tile with successful match. Notice that the number of stars with the initial selection is presented here (magnitudes $=[14,21]$ and errors $<0.1$). 

\begin{table}
	\caption{Average number of stars used per field. We list the typical number of matches to the other catalogs; the variations reflect different depths and field sizes. When matching ATLAS to SDSS, many fields have few matching stars because they are on the edges of SDSS, thus we adopt a larger tile size than KiDS to obtain more matched stars per field. The NSC sample is a combination of many shallow exposures and several deep exposures, so the average number of matched stars is also lower than KiDS.}
	\label{tab:star_num}
	\centering
	\begin{tabular}{|r|c|c|c|}
		\hline
	    & KiDS  & ATLAS & NSC \\ \hline
	    tile size & $1^{\circ}\times1^{\circ}$ & $2^{\circ}\times2^{\circ}$ & $1.5^{\circ}\times1.5^{\circ}$ \\
        minimum no. of stars detected & $\sim$200  & $\sim$1000  & $\sim$1100   \\
        maximum no. of stars detected & $\sim$1800 & $\sim$14000 & $\sim$13800 \\
	    average no. of stars detected & $\sim$800  & $\sim$3700  & $\sim$3000  \\
		matched with SDSS             & $\sim$700  & $\sim$1400  & $\sim$600   \\
		matched with Refcat2          & $\sim$600  & $\sim$3400  & $\sim$560   \\
		matched with PanSTARRS        & $\sim$700  & $\sim$3500  & $\sim$570   \\
        \hline
    \end{tabular}
\end{table}

\subsection{Color Equations}
In order to test our method, we will compare (re-calibrated) \textit{u}- and \textit{g}-band magnitudes to SDSS and PanSTARRS.  Since the filter transmission curves from different instruments/surveys are not identical, this invariably requires transforming one set of magnitudes into another using color-dependent equations.
For KiDS and ATLAS, such transformations (based on cross-comparison to SDSS) have been published in the literature, but these do not suffice for our purposes:  for KiDS, the color equations are reported only after subtracting an unspecified constant \citep{kids_data}; for ATLAS, the \textit{u}-band transform to SDSS is specified as a linear color equation, but it is noted that a second-order fit would be preferable \citep{atlas_web}. We initially attempt to re-derive the color transformations from overlapping survey data (App.~\ref{sec:appb}); however, we find systematic offsets between KiDS and ATLAS {\it u}- and {\it g}-band magnitudes, despite stemming from the same telescope and instrument.  Similarly, we compare the NSC {\it u}- and {\it g}-band magnitudes with those published by the SHELA survey \citep{shela}; note that the SHELA data is also included in the NSC - with an independent data reduction - and therefore does not constitute an additional validation dataset for our main purpose), and find systematic differences.  We therefore choose to not use color transformations from samples of cross-matched stars between KiDS/ATLAS/NSC and SDSS/PanSTARRS, since that would make our method dependent on assuming that one of these calibrations is correct and the other is not.
For completeness, we include our own measurements of the color transforms from the cross-matched survey data in App.~\ref{sec:appb}. We note that these color transformations show discrepancies with the color transformations derived from synthetic magnitudes (see below), but a full investigation of these discrepancies (shown in Fig.~\ref{fig:synth}) is beyond the scope and purpose of this work.

Rather than relying on the calibration of one of the validation datasets, we opt to to fit the color equations based on synthetic magnitudes, following the SDSS \citep{synth1, synth2} and PanSTARRS \citep{pan_synth1, pan_synth2} procedures.  Synthetic magnitudes can be derived by convolving the spectra of well-calibrated spectrophotometric standard stars with the filter transmission functions of the instruments. For each star, the synthetic magnitudes yield the
colors (i.e. magnitude differences, i.e. flux ratios) between different magnitudes, including between those from different instruments.  By using a library of spectrophotometric standard stars that representatively samples the stellar locus, these relations can be determined as function of a specific color, i.e. the color transformation equations. 
The synthetic color equations are independent of the initial calibration of the validation sets, and are able to link the calibration accuracy to an absolute flux standard, in this case the AB system of SDSS. Because of their importance to photometric calibration, significant investments have been made to accurately and precisely map the spectra of sets of spectrophotometric standard stars, which puts us in a position to capitalize on this work.  
We therefore calculate synthetic magnitudes for all instruments involved in this work using stellar spectra from Pickles \citep{pic}, XSL \citep[The X-Shooter Spectral Library,][]{xsl_dr1, xsl_dr2} and CALSPEC \citep{cal_rev, calspec}. We present the color transform equations and their determination in App.~\ref{sec:appc}.

\section{Method Construction}
\label{sec:model}

\begin{figure}
    \includegraphics[width=\columnwidth]{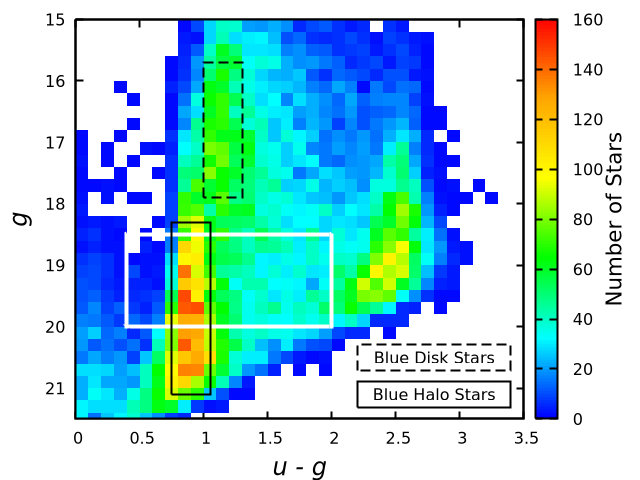}
	\caption{2d histogram of stars in a 4$\times$4 deg$^2$ SDSS field of \textit{g}-band magnitude and \textit{u-g} color. The solid black box indicates the population dominated by halo MSTO stars, and the dashed black box indicates blue disk stars. Stars within the white box are preselected to be used for detecting the edge \textit{u-g} color (see Sect.~\ref{subsec:iso} and Fig.~\ref{fig:edge_model}). The \textit{g} magnitude cut at 20.0 is chosen for a relatively complete sample selection. The \textit{u-g} pre-selection can be set by eye to include stars on both sides of the blue halo stars, and can be different for different surveys and un-calibrated instrumental magnitudes.}
    \label{fig:iso}
\end{figure}

We here construct a model for the correlation between the \textit{u-g} color of the ``blue tip'' stars and their galactic latitude.
\citet{Ivezic} showed that there are two distinct populations of MSTO stars in the Milky Way: one with a brighter magnitude ($g\lesssim17$) and a redder \textit{u-g} color, while the other are fainter and bluer. Within each population, the \textit{u-g} color (a proxy for stellar metallicity) is nearly constant regardless of their magnitudes.
The brighter, redder population stems from MSTOs in the thick galactic disk, while the fainter, bluer population is dominated by halo MSTOs which are older and have a lower metallicity. In Fig.~\ref{fig:iso} we replicate this finding. \citet{Ivezic} find that the metallicity of the halo population is spatially invariant, as is also shown by other studies \citep[e.g.][]{ibata17}.  This is the key assumption that we will utilize here.

We first isolate likely halo MSTO stars from the distribution of \textit{u-g} colors (Fig.~\ref{fig:iso}). The first step is to isolate the population of stars approximately within the white box shown in Fig.~\ref{fig:iso}, including stars on both sides of halo MSTOs while excluding the population of red stars.
For un-calibrated data, the specific magnitude selection will be different according to the instrumental zero-point.  We then fit a step-function-like model to find the edge color by counting the stars on both sides of the edge, described in Sect.~\ref{subsec:iso}. In SDSS data we find a dependence of the detected edge color on galactic latitude, which we attribute to contamination from disk MSTOs along the line of sight. We fit the color-latitude dependence with a plane-parallel disk model in Sect.~\ref{subsec:con}. We also test the impact of spatial inhomogeneities in the halo on the relation by examining areas near the Sagittarius Stream, the largest stellar structure in the galactic halo. We find that any effect is too small for us to detect (Sect.~\ref{subsec:sag}). In Sect.~\ref{sec:rec}, we show how the color-latitude relation can be used to calibrate \textit{u}-band photometry, and verify the method on KiDS, ATLAS and NSC data sets.

\subsection{Detecting the ``Blue Tip''}
\label{subsec:iso}

\begin{figure*}
	\centering
    \includegraphics[width=\columnwidth]{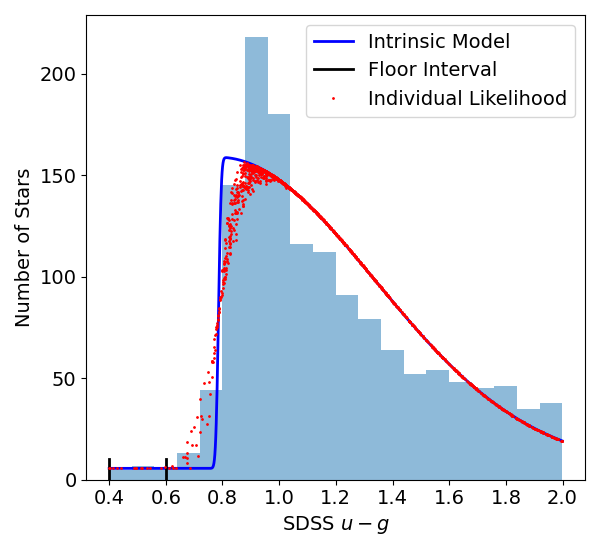}
    \includegraphics[width=\columnwidth]{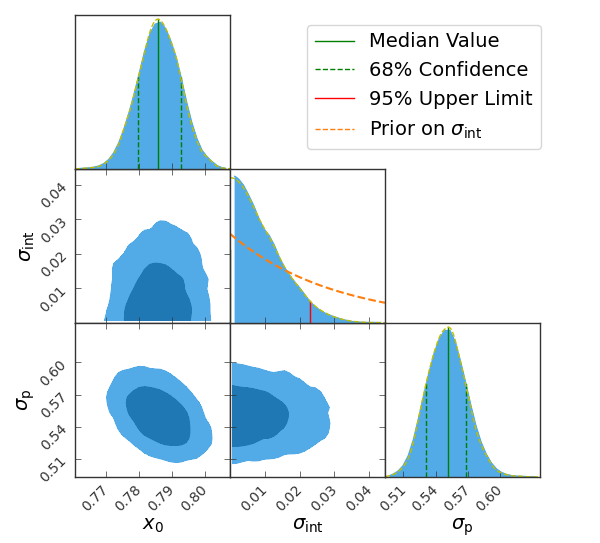}
	\caption{Left: histogram of the \textit{u-g} colors of stars from one SDSS field (within the pre-selected magnitude and color range) and the intrinsic color distribution model (blue line; Eq.~\ref{eqn:intrinsic_model}). 
	The individual likelihood (Eq.~\ref{eqn:conv}) of each star is plotted in red points; these convolve the intrinsic model with the measurement uncertainty, effectively broadening the blue edge. Both the intrinsic model and the invididual likelihood are re-scaled to match the histogram. Notice that the model is not fit to the histogram, but by evaluating the likelihood for each individual star (Eq.~\ref{eqn:conv}). The parameters used for plotting both models are taken from the best-fit convolved model (Eq.~\ref{eqn:conv}).  Black vertical lines mark the Floor Interval for this field, within which the floor probability $F$ is derived. Right: Posterior distribution of the model parameters. Notice that the intrinsic edge width $\sigma_{\mathrm{int}}$ is consistent with zero, affirming the assumption that it is small. The figures are created with functions defined in this \href{https://github.com/shuang92/blue_tip_calibration/blob/main/edge_detection_emcee.ipynb}{notebook}.}
    \label{fig:edge_model}
\end{figure*}

The goal is to detect the blue-side edge of the MSTOs' \textit{u-g} color, as shown in the left panel of Fig.~\ref{fig:edge_model} where the \textit{u-g} distribution of MSTOs in an SDSS field is plotted.
The stars included in Fig.~\ref{fig:edge_model} are pre-selected in \textit{g}-band magnitude and \textit{u-g} color as discussed previously (the white box in Fig.~\ref{fig:iso}). Note that the distribution extends to the redder side of the blue edge by about one magnitude, to help identify the blue edge. The pre-selection cuts are different for each survey and they are listed in Table~\ref{tab:FI}.
\begin{table}
	\caption{Data selection for the ``blue tip'' model. The \textit{g} magnitude cuts and the \textit{u-g} pre-selection define the white box in Fig.~\ref{fig:iso}, inside which the stars are used for detecting the edge color. ``Floor Interval'' specifies the {\it u-g} range used to derive the floor probability density $F$ in the model (equation [\ref{eqn:intrinsic_model}]) and [\ref{eqn:conv}]). The magnitude cut at 20.0 is chosen to compensate the completeness of some shallow fields in NSC. Color cuts and Floor Intervals are determined by eye examination and need to be adjusted changes in a few irregular fields.}
	\label{tab:FI}
	\centering
	\begin{tabular}{|r|c||c|c}
		\hline
		       & mag \textit{g} cut & \textit{u-g} pre-selection ($a-b$) & Floor Interval  \\ \hline
		SDSS   &  $18.5 - 20.0$ & $0.4 - 2.0$ & $0.4 - 0.6$  \\
		KiDS   &  $18.5 - 20.0$ & $0.0 - 2.0$ & $0.0 - 0.6$  \\
		ATLAS  &  $18.5 - 20.0$ & $0.0 - 2.0$ & $0.0 - 0.6$  \\
		NSC    &  $18.5 - 20.0$ & $0.5 - 2.0$ & $0.5 - 0.8$  \\
        \hline
    \end{tabular}
\end{table}

Our starting point is the ``blue tip'' model of \citet{blue-tip}:
\begin{equation}
   \qquad\qquad\qquad P(x|x_0) = \frac{1}{2} \left[ 1 + \mathrm{Erf} \left(\frac{x-x_0}{\sqrt{2}\sigma_{\mathrm{int}}} \right) \right] + F.
\end{equation}
It is a probability distribution of the \textit{u-g} color (denoted $x$) of each star in the field given the MSTO edge color $x_0$. Here Erf() is the Gaussian Error function, the result of a step function convolved with a Gaussian of width $\sigma_{\mathrm{int}}$. In \citet{blue-tip}, the only free parameter in the model is the MSTO edge color $x_0$, while the intrinsic width of the edge $\sigma_{\mathrm{int}}$ and the floor probability $F$ (number of stars per magnitude on the left side of the edge) are set to fixed values.

We make several modifications to the model:
\begin{enumerate}
    \item[(i)] We multiply by a Gaussian PDF centered on $x_0$ with width $\sigma_p$ to better describe the distribution on the right side of the edge.
    \item[(ii)] We make $\sigma_{\rm int}$ a free parameter in the fit.
    \item[(iii)] The floor probability density $F$ is calculated by counting the actual number of stars on the left side of the edge.
\end{enumerate}
The model thus becomes:
\begin{equation}
    \quad P(x|x_0, \sigma_p, \sigma_{\mathrm{int}}) = \frac{A}{2}\mathcal{N}(x_0, \sigma_p)
    \left[ 1 + \mathrm{Erf}\left( \frac{x-x_0}{\sqrt{2}\sigma_{\mathrm{int}}} \right) \right] + F,
    \label{eqn:intrinsic_model}
\end{equation}
where $A$ is the normalization factor for $P(x)$ to integrate to 1 over the sample range. Given the {\it u-g} bounds $a$ and $b$ of the pre-selection, $A$
is determined from other model parameters: the edge color $x_0$, the width of the right-side Gaussian $\sigma_p$, the floor probability $F$, as well as $a$ and $b$. Assuming that $\sigma_\mathrm{int} \ll \sigma_p$, $A$ can be calculated from:
\begin{equation}
    \qquad\qquad A \cdot \mathrm{Erf} \left( \frac{b-x_0}{\sqrt{2}\sigma_\mathrm{p}} \right) = 2 - 2F(b-a).
    \label{eqn:A}
\end{equation}
Note that the model is robust against the arbitrariness of $a$ and $b$ (they are decided by eyes, see Table~\ref{tab:FI}), since any small changes in $a$ and $b$ contribute only to the global normalization $A$ and thus do not alter the probability of one star's {\it u-g} color relative to another. Similarly, the floor probability density $F$, like $a$ and $b$, has some arbitrariness in its determination. It is calculated by counting stars inside a pre-defined Floor Interval (FI) of the \textit{u-g} color,
\begin{equation}
   \qquad\qquad F= \left( \frac{N^{\mathrm{FI}}}{N} \right) / \mathrm{len(FI)},
   \label{eqn:F}
\end{equation}
where $N^{\mathrm{FI}}$, $N$ and len(FI) are the number of stars in the Floor Interval, the number of stars in the pre-selected sample (i.e. the white box in Fig.~\ref{fig:iso} and the length of the Floor Interval, respectively. Our choices for Floor Intervals for all surveys are summarized in Table~\ref{tab:FI}.

We refer to model (\ref{eqn:intrinsic_model}) as the intrinsic model. We make a further modification to account for the photometric uncertainties:
\begin{enumerate}
    \item[(iv)] For each star, the intrinsic model is convolved with the observed (photometric) uncertainty of the \textit{u-g} color.
\end{enumerate}
The convolution calculation is challenging. However, it can be substantially simplified with the assumption that the intrinsic edge width $\sigma_{\mathrm{int}}$ is small compared to both the width of the right-side Gaussian ($\sigma_\mathrm{P}$) and the \textit{u-g} photometry uncertainties (as we will see, $\sigma_{\mathrm{int}}$ is often consistent with zero, so this assumption holds true). Under this assumption, the convolved model (which now represents the individual likelihood of each star) becomes
\begin{equation}
    \mathcal{L}_i^{\mathrm{star}}(x_i|x_0, \sigma_p, \sigma_{\mathrm{int}}) = \frac{A}{2}\mathcal{N}(x_0, \sigma_p)
    \left[ 1 + \mathrm{Erf}\left( \frac{x_i-x_0}{\sqrt{2}\Sigma_i} \right) \right] + F,
    \label{eqn:conv}
\end{equation}
where $i$ labels each single star, and $\Sigma_i^2 = \sigma_\mathrm{int}^2 + u_{\mathrm{err},i}^2 + g_{\mathrm{err},i}^2$ is the square sum of the intrinsic edge width and the photometric uncertainty of the {\it u} and {\it g} magnitudes. Comparing Eq.~\ref{eqn:conv} to Eq.~\ref{eqn:intrinsic_model}, we see that the convolution with the photometric uncertainty results in a broadening of the error function (i.e. the ``step''), as one would have intuitively guessed.
An example of the intrinsic and convolved models are shown in the left panel of Fig.~\ref{fig:edge_model} on top of the stars to which they are fit.

In sum, our model has 3 parameters: the edge color $x_0$, the intrinsic edge width $\sigma_{\mathrm{int}}$ and the width of the right side Gaussian $\sigma_\mathrm{p}$. $A$ and $F$ are not free parameters and are calculated with Eq.~(\ref{eqn:A}) and Eq.~(\ref{eqn:F}) respectively. We use flat priors on $x_0$ and $\sigma_\mathrm{P}$, and an exponential prior $e^{-3\sigma_{\mathrm{int}}}$ on $\sigma_{\mathrm{int}}$ to reflect our assumption that it is small and to ensure the convergence of the chains. We use PyStan \citep{pystan}, a Python wrapper of Stan\footnote{\url{https://mc-stan.org/users/documentation/}} to perform an MCMC sampling. The posteriors of one field are shown in the right panel of Fig.~\ref{fig:edge_model}. In most cases, we can only constrain an upper limit on $\sigma_{\mathrm{int}}$ which is indeed much smaller than $\sigma_\mathrm{p}$ or the \textit{u-g} uncertainty, typically 0.5 mags and 0.05 mags respectively.  

\subsection{Disk Contamination Model}
\label{subsec:con}

\subsubsection{The Color-Latitude Model}

\begin{figure*}
	\includegraphics[width=\columnwidth]{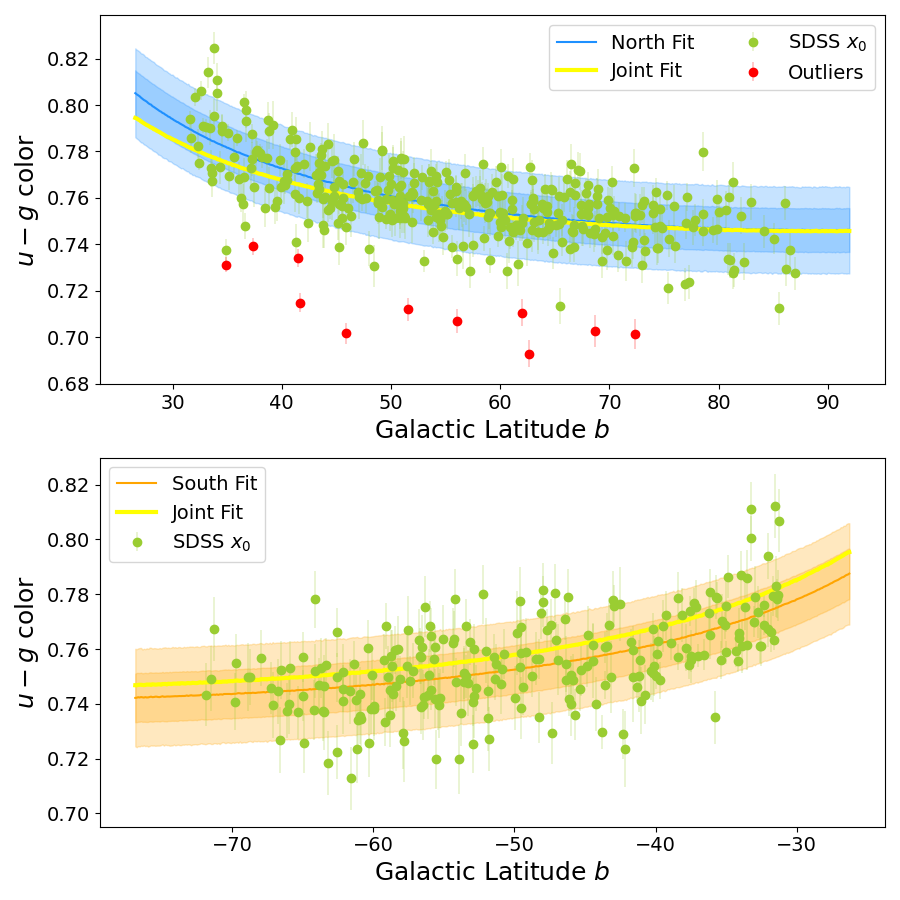}
	\includegraphics[width=\columnwidth]{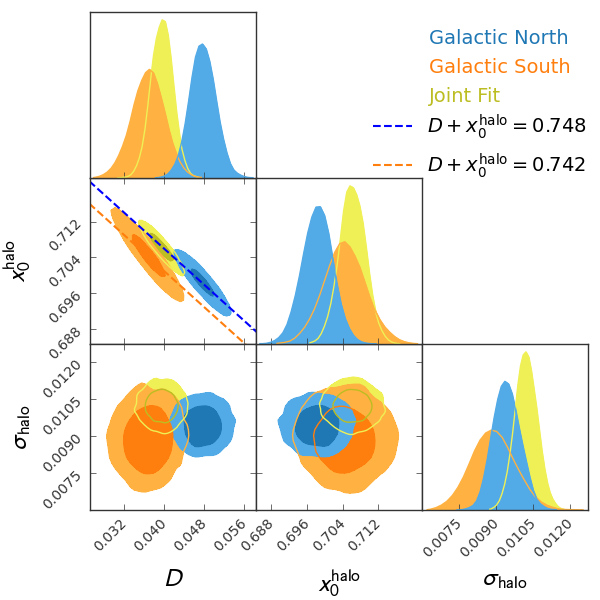}
	\caption{The correlation between the \textit{u-g} edge color of SDSS halo MSTOs ($x_0$) and galactic latitude. Left: SDSS data and best-fit disk contamination models (equation [\ref{equ:fit}]). Each data point is the measured \textit{u-g} edge color of MSTOs, $x_0$, in a field. The separate fits in the Galactic North and South are shown with blue and orange colors respectively, with the shaded areas being $1\sigma$ and $2\sigma$ width of all curves generated from MCMC samples of ($D, x_0^{\mathrm{halo}}$) and perturbed with a random Gaussian noise of width $\sigma_{\mathrm{halo}}$, the intrinsic scatter of model (equation [\ref{equ:fit}]). The best-fit joint model is plotted in yellow. Some outliers in the North are marked in red; these are exposures taken in two specific scans, run 4674 and 4678 (see Sect.~\ref{subsec:fit}) and are excluded from the fitting process due to their inconsistent photometry.  Right: Posterior probability distribution of free parameters $D$, $x_0^{\mathrm{halo}}$ and $\sigma_\mathrm{halo}$ of the model. The dashed lines indicate constant $D + x_0$ values of the best-fit $D$ and $x_0^{\mathrm{halo}}$ from the separate fits (see Table~\ref{tab:sdss}). These $D + x_0^{\mathrm{halo}}$ values represent the observed MSTO colors with minimum contamination at $b=\pm90^\circ$. The figures are created with this \href{https://github.com/shuang92/blue_tip_calibration/blob/main/disk_model_emcee.ipynb}{notebook}.}
	\label{fig:pos}
\end{figure*}

The procedures described in Sect.~\ref{subsec:iso} deliver a robust edge \textit{u-g} color ($x_0$ in the model) for the halo MSTOs in each field. \citet{blue-tip} find a slow spatial variation of the edge color across the sky, which they attribute to increasing contamination of the sample by redder disk stars.  Following this interpretation, we show the measured edge color as function of galactic latitude $b$ in both the North and the South in  Fig.~\ref{fig:pos}, and find a notable correlation.

Since dust extinction has already been accounted for, we follow \citeauthor{blue-tip} and attribute this correlation to the contamination of disk stars in our halo MSTO sample. The contamination is unavoidable since the line of sight always goes through the galactic disk to reach the halo. Some of the disk stars, more distant and thus fainter ($g>18.5$), may enter our sample selection, causing the blue tip to shift redder. For simplicity, we here assume a flat disk structure of the Milky Way to model the contamination - as we will see, the remaining scatter around this simple model matches the precision of the SDSS {\it u}-band calibration, indicating that a more complex model is not necessary.
Then, the number of disk stars in each exposure, and thus the amount of contamination, is roughly proportional to the path length through the disk. We model this contamination with a simple plane-parallel model (similar to airmass):
\begin{equation}
	\qquad\qquad\qquad\qquad   \widehat{x_0} = \frac{D}{|\sin{b}|} + x_0^{\mathrm{halo}}
	\label{equ:fit}
\end{equation}
where $\widehat{x_0}$ is the model prediction of the observed (contaminated) edge \textit{u-g} color, $x_0^{\mathrm{halo}}$ the intrinsic (uncontaminated) edge color of the pure halo MSTOs and $D$ the disk contamination factor. The denominator $|\sin{b}|$ represents the line-of-sight length within the galactic disk, with $b$ being the galactic latitude of the field. The detected edge color $x_0$ often has a Gaussian-like posterior distribution (Fig.~\ref{fig:edge_model}), thus we model the measurements with a Gaussian likelihood:
\begin{equation}
	\mathcal{L}_i^{\mathrm{field}}(x_0^i|D, x_0^{\mathrm{halo}}, \sigma_{\mathrm{halo}}) = \frac{1}{\sqrt{2 \pi (\sigma_i^2 + \sigma_{\mathrm{halo}}^2})} 
	\exp\left[ -\frac{(x_0^i - \widehat{x_0})^2}{2 (\sigma_i^2 + \sigma_{\mathrm{halo}}^2)} \right] ,
	\label{equ:L}
\end{equation}
where $i$ stands for each field, $x_0^i$ is the detected edge color of the halo MSTOs in the field ($x_0$ in equation [\ref{eqn:conv}]), 
$\sigma_i$ is the uncertainty on $x_0$ (the $1\sigma$ width of $x_0$ as shown in the right panel of Fig.~\ref{fig:edge_model}) and $\sigma_{\mathrm{halo}}$ is the intrinsic scatter of the color-latitude relation.

\subsubsection{Fitting Results}
\label{subsec:fit}

We fit the parameters $D$, $x_0^{\mathrm{halo}}$ and $\sigma_{\mathrm{halo}}$ with flat priors. To avoid using SDSS for both the calibration and the verification, we fit the model using only the regions outside of the validation sets. 

Since the galactic halo is known to be much more complicated than a single-component sphere \citep{merge, splash, simu}, we initially fit the Galactic North ($b>0$) and South ($b<0$) separately. Furthermore, \citet{blue-tip} find a color asymmetry between the North and South (see table~4 therein). 

We again use PyStan to fit the model with an MCMC and show the results in Fig.~\ref{fig:pos} and Table~\ref{tab:sdss}. We find some outliers in the Galactic North which are attributable to exposures taken in two specific SDSS runs\footnote{\url{https://www.sdss.org/dr16/imaging/imaging_basics/}} (a contiguous period of drift scan), 4674 and 4678 (Fig.~\ref{fig:runs}). We exclude those tiles from the MCMC fit, assuming that they suffer from systematic calibration issue \citep[see also fig.~6 of ][]{blue-tip}. 

We find marginal evidence for (small) differences in the color-latitude dependence between the North and South, since the $D$-$x_0^{\mathrm{halo}}$ posteriors do not overlap in the two-dimensional view (Fig.~\ref{fig:pos}).  $D$ and $x_0^{\mathrm{halo}}$ are highly degenerate, and the data best constrains the combination ($D+x_0^{\mathrm{halo}}$), corresponding to the blue edge color at $|b|=90^\circ$ where the disk contamination is minimal. The fact that the $D$-$x_0^{\mathrm{halo}}$ posteriors do not overlap can be attributed to a slight mismatch between the blue edge colors at $|b|=90^\circ$ in the best-fit model. The color difference (6 mmags) is in the opposite direction to \citet{blue-tip}, where the North is bluer than the South by 7.6 mmags (comparing the median colors of $40^\circ < |b| < 70^\circ$, see table~4 therein). Moreover, we note that this difference is smaller than the measured intrinsic scatter of the relation ($\sim$9 mmag). We also present a joint fit both the North and South in Fig.~\ref{fig:pos} and Table~\ref{tab:sdss}.  The joint fit absorbs the small North-South difference into a slightly increased intrinsic scatter. The intrinsic scatter from the joint fit, $10.2 \pm 0.4$ mmags, is still smaller than the SDSS residual calibration uncertainty (per run) on the \textit{u}-band \citep[15 mmags;][]{hypercal}; this is most likely due to a different tile size: in this work, we use $2\times2$ deg$^2$ fields from SDSS, while an SDSS run consists of 6 columns, each $13'.5$ wide.  Since the inferred North-South difference is quite small, and smaller than the best-fit intrinsic scatter, we continue with the model from the joint fit, describing both hemispheres with the same model parameters, for simplicity.
\begin{table*}
	\centering
	\caption{Maximum {\it a posteriori} parameters for SDSS disk contamination model (equation [\ref{equ:fit}]) with $1\sigma$ uncertainties. Left: fitting all SDSS stars regardless of being in the Sagittarius Stream or not. The North and South fit yields consistent marginalized parameters. The joint fit is used for the \textit{u}-band calibration. Right: Separating the field stars from the Sagittarius Stream stars. The stream stars are allowed to have a different intrinsic color than that of the field stars. The fitting yields no significant difference between the two.} 
	\label{tab:sdss}
	\begin{tabular}{cccc c cccc}
\hline
               & $D$  &$x_0^{\mathrm{halo}}$ & $\sigma_\mathrm{halo}$ & 
        \vline & $D$  & $x_\mathrm{0,field}^{\mathrm{halo}}$ & $x_\mathrm{0,stream}^{\mathrm{halo}}$  & $\sigma_\mathrm{halo}$  \\ \hline 

Galactic North & $0.048 \pm 0.002$ & $0.700 \pm 0.003$ & $0.0094 \pm 0.0006$ & 
        \vline & $0.046 \pm 0.003$ & $0.702 \pm 0.004$ & $0.695 \pm 0.004$ & $0.0091 \pm 0.0005$ \\

Galactic South & $0.037 \pm 0.003$ & $0.705 \pm 0.005$ & $0.0088 \pm 0.0009$ & 
        \vline & $0.038 \pm 0.004$ & $0.703 \pm 0.005$ & $0.705 \pm 0.005$ & $0.0090  \pm 0.0009$   \\

Joint fit      & $0.040 \pm 0.002$ & $0.705 \pm 0.003$ & $ 0.0102 \pm 0.0005$ & 
        \vline & $0.038 \pm 0.002$ & $0.709 \pm 0.003$ & $ 0.705  \pm 0.003$ & $0.0101 \pm 0.0005$  \\
\hline
    \end{tabular}%
\end{table*}

\subsubsection{The Sagittarius Stream}
\label{subsec:sag}

The Sagittarius Stream is the most dominant structure in the galactic halo, as shown in Fig.~\ref{fig:cover}. We test whether fields within the Sagittarius Stream have a notably different {\it u-g} edge color by adding an additional free parameter, $x_\mathrm{0,stream}^{\mathrm{halo}}$, to the model. We keep $D$ and $\sigma_{\mathrm{halo}}$ the same for tiles in and out of the stream, while allowing their halo colors to be different, parameterized with $x_\mathrm{0,field}^{\mathrm{halo}}$ for the field tiles, and $x_\mathrm{0,stream}^{\mathrm{halo}}$ for the tiles inside the Sagittarius Stream. All tiles in and out of the stream are fit simultaneously with the 4 parameters, using either the North/South separate model or the joint model. The stream tiles are defined to be within $7^\circ$ from zero latitude of the Sagittarius Coordinate System \citep{Sag}. In all models, the stream tiles show no significant color difference from the field tiles, as shown in Table~\ref{tab:sdss}. We suspect that this is because most stars in the Sagittarius Stream are too faint for our sample selection; Fig. 25 of \citet{sesar10} indicates that stars belonging to the Stream are predominantly at $20 \lesssim r \lesssim 21.5$, whereas we select stars with $18.5 < g < 20$ and typically blue color $g-r \sim 0.4$.

\section{Re-calibration of the Validation Sets}
\label{sec:rec}

The key step to using the {\it u}-band calibration method for a specific data set is to determine the parameters $D$ and $x_{0}^{\rm halo}$ of the color-latitude relation equation (\ref{equ:fit}) for the instrument being used. Once these parameters, and thus the relation, are established, any new data set taken with the same instrument can be calibrated by direct comparison between the measured and predicted blue tip \textit{u-g} color $x_0$. As the filter responses and CCD efficiency of a given instrument are generally different from SDSS, a different blue tip color $x_0$ will be measured. We here assume that the contamination factor $D$ is universal among instruments with SDSS-like filters, since it is determined by the geometry of the Milky Way. The instrument-specific halo color $x_0^{\mathrm{halo}}$ can be derived by minimizing the \textit{u}-band residual with respect to SDSS on overlapping fields. We demonstrate this process for VST (the KiDS and ATLAS data sets) and DECam (the NSC data) in Sect.~\ref{subsec:calib}.

As the calibration is based on the \textit{u-g} color, a well-calibrated \textit{g}-band is necessary. Since we find inconsistencies between published color transformations from different surveys conducted with the same instruments (App.~\ref{sec:appb}), we choose to first re-calibrate the \textit{g}-band survey data (Sect.~\ref{subsec:gdis}).

\subsection{The \textit{g}-band calibration}
\label{subsec:gdis}

Each survey considered here follows a different methodology for calibrating {\it g} photometry:  KiDS first uses the stellar locus to find the $(g-r)$ zeropoint, and then calibrates {\it r}-band to Gaia DR2 with a color equation calibrated in the SDSS-KiDS overlap \citep{kids}.
Earlier ATLAS data releases calibrated directly to APASS, the AAVSO (American Association of Variable Star Observers) Photometric All-Sky Survey\footnote{\url{https://www.aavso.org/apass}} \citep{atlas, apass} for {\it g}-band; in DR4, this was superseded by calibrating directly to Gaia DR2, using a color equation between VST {\it g}-band and Gaia photometry \citep{atlas_web}.
NSC calibrates to PanSTARRS1 in the North, and derives an extinction-dependent color equation between the DECam {g}-band and the APASS {\it g}-band, using 2MASS photometry for the color dependence \citep[][note that we here can only compare to the PanSTARRS-calibrated sky area]{nsc}.

\begin{figure*}
        \includegraphics[width=0.6\columnwidth]{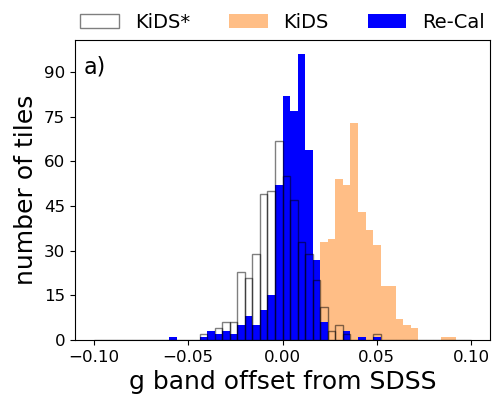} 
        \includegraphics[width=0.6\columnwidth]{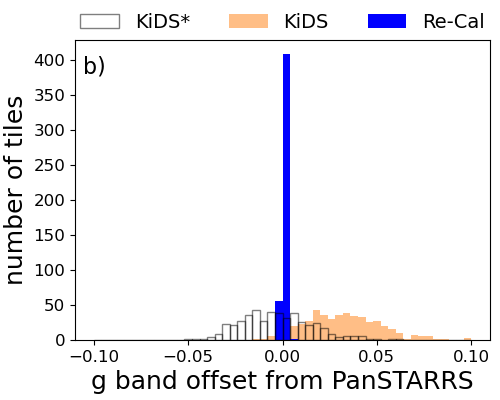}
        \includegraphics[width=0.6\columnwidth]{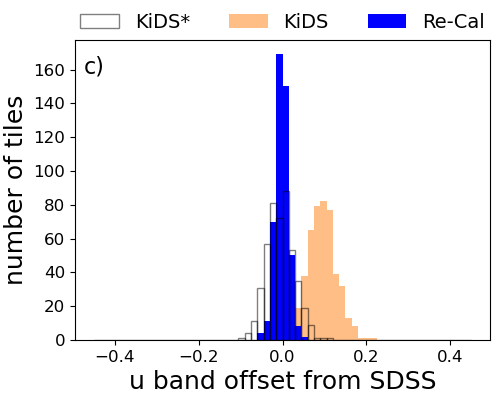}
        
        \includegraphics[width=0.6\columnwidth]{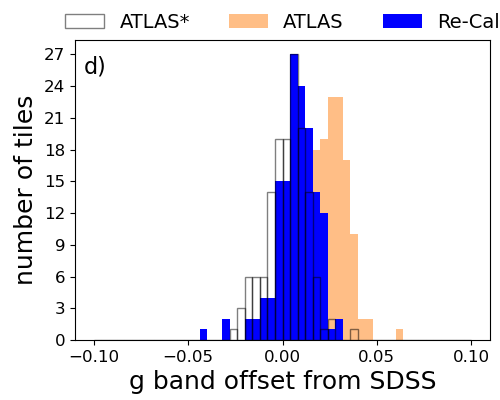}
        \includegraphics[width=0.6\columnwidth]{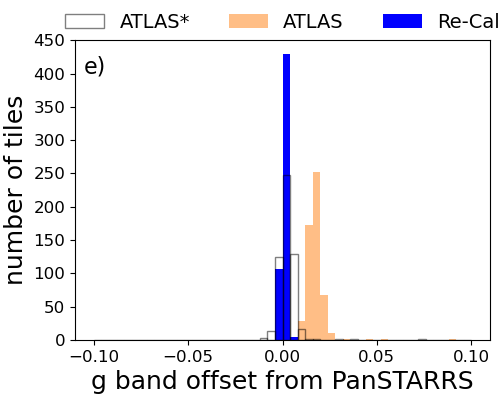}
        \includegraphics[width=0.6\columnwidth]{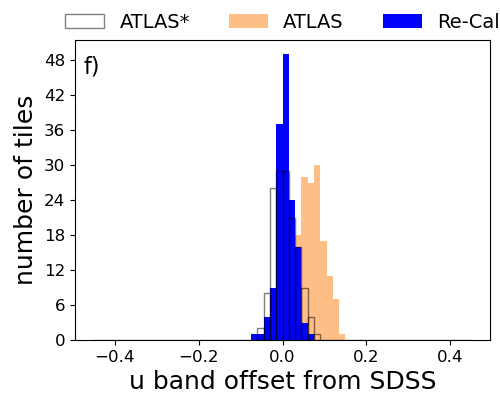}
            	
        \includegraphics[width=0.6\columnwidth]{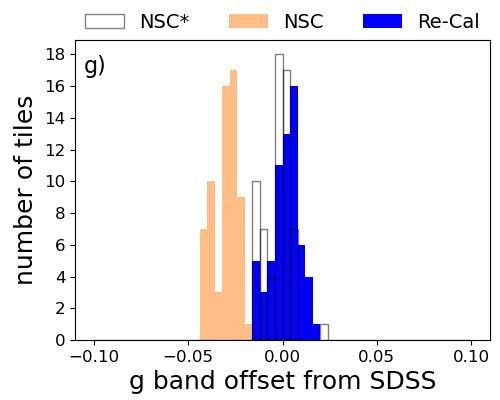}
    	\includegraphics[width=0.6\columnwidth]{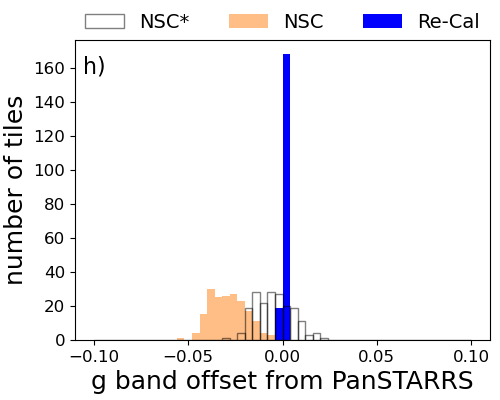}
    	\includegraphics[width=0.6\columnwidth]{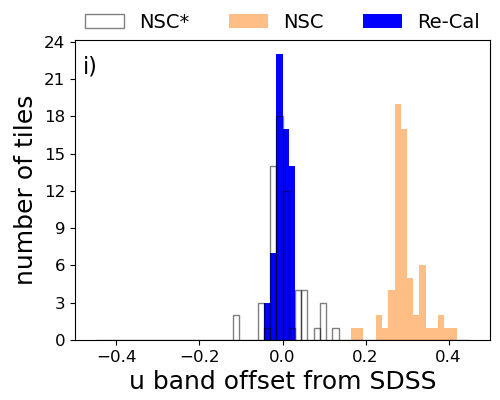}

        \caption{The distributions of the median magnitude differences (computed per field) of the validation sets compared with reference truth catalogs. The orange histograms represent the original data release of KiDS/ATLAS/NSC, comparing to the reference data with color equations fit on the synthetic magnitudes (equation [\ref{eqn:color_synth}]). The blue histograms are the offsets after our re-calibration, also comparing to the references with equation (\ref{eqn:color_synth}). For comparison, the empty histograms show the offsets from the original data release, but with color equations fit on the cross-matched data sets (equation  [\ref{eqn:color}]). The empty histograms thus center on zero mean values by construction. The standard deviation of the original (orange) and re-calibrated (blue) distributions are listed in Table~\ref{tab:res}. Note that because the {\it g}-band re-calibration calibrates with respect to Refcat-2, which is heavily dominated by PanSTARRS, the re-calibrated {\it g}-band is in excellent agreement with PanSTARRS by construction.}
        \label{fig:res}
\end{figure*}

We choose to re-calibrate the \textit{g}-band zero-points based entirely on Refcat2, due to its all-sky photometric homogeneity.  We will compare the results with SDSS, as well as PanSTARRS for a larger sample size beyond the SDSS footprint. The \textit{g}-band calibration follows the procedures:
\begin{enumerate}
	\item Match stars between Refcat2 and the validation sets as described in Sect.~\ref{subsec:match}.
	\item Convert Refcat2 \textit{g} to $g_{_{\mathrm{VST/DECam}}}$ using equation (\ref{eqn:color_synth}) and calculate the \textit{g}-band offset for each matched star.
    \item Calculate the median offset for each field.
	\item Adjust the {\it g}-band zeropoint by subtracting the median in each field, to yield calibrated {\it g}-band magnitudes.
\end{enumerate}
The verification with SDSS/PanSTARRS follows the same routine:
\begin{enumerate}
	\item Match stars between SDSS/PanSTARRS and the validation sets.
	\item Convert $g_{\mathrm{SDSS/Pan}}$ to $g_{_{\mathrm{VST/DECam}}}$ using equation (\ref{eqn:color_synth}).
	\item Calculate the median-tile \textit{g}-band offsets, before and after the re-calibration.
\end{enumerate}
The resulting median-tile \textit{g}-band offsets are shown in the left and middle columns of Fig.~\ref{fig:res}, with orange/blue color representing the original/re-calibrated data sets. We compare the measured standard deviations of the offset distributions in Table~\ref{tab:res}.
For comparison, in Fig.~\ref{fig:res} we also show (as empty histograms) the magnitude offsets between the original photometry and the validation datasets when transforming magnitudes with color equations fit directly on the cross-matched data sets (equation [\ref{eqn:color}]).

\begin{table*}
	\centering
	\caption{The median tile-offset standard deviation of the validation set comparing with other surveys, before and after our re-calibration. This table corresponds to Fig.~\ref{fig:res}, where the original offsets are shown in empty histograms and the calibrated offsets in blue. The standard deviations and the uncertainties in this table are derived from a boot-strap re-sampling from the corresponding distributions in Fig.~\ref{fig:res}.}
	\begin{tabular}{lccc}
	\hline
                 &\multicolumn{2}{c}{$g$-band std, compared to:} &  $u$-band std to SDSS       \\
                 &           SDSS        &       PanSTARRS       & (joint $ug$ calibration)    \\ \hline
KiDS original    &  $0.0134 \pm 0.0005$  & $0.0191\pm0.0007$     &   $0.033 \pm 0.002$         \\
KiDS calibrated  &  $0.0110 \pm 0.0007$  & $0.00108\pm0.00004$   &   $0.0157\pm0.0006$         \\ \hline
ATLAS original   &  $0.0106 \pm 0.0006$  & $0.005 \pm 0.001$     &   $0.027 \pm 0.001$         \\
ATLAS calibrated &  $ 0.011 \pm 0.001$   & $0.00094 \pm 0.00005$ &   $0.020 \pm 0.002$         \\ \hline
NSC original     &  $0.0069 \pm 0.0006$  & $0.0096\pm0.0004$     &   $0.041 \pm 0.005$         \\
NSC calibrated   &  $0.0074 \pm 0.0006$  & $0.00077\pm0.00006$   &   $0.016 \pm 0.001$         \\ \hline
    \end{tabular}
    \label{tab:res}
\end{table*}

Notice that the empty histograms (the original magnitudes converted into the SDSS system with color equations fit from overlapping data) are centered on zero by construction, since the constant offsets are absorbed into the color equations. The orange histograms (those converted with color equations from synthetic magnitudes), on the other hand, reveal a systematic difference between the validation sets and the SDSS AB system. After the re-calibration to Refcat-2, the orange histograms show negligible mean values with respect to both SDSS and PanSTARRS, indicating an improvement in systematic photometric accuracy.

Also notice that, after calibration, the RMS scatters w.r.t. PanSTARRS reduce to $\le 1$ mmag level. This is expected, since PanSTARRS is the major component of Refcat2 within its footprint. 
For the offsets w.r.t. SDSS, we find similar scatters before/after re-calibration on all data sets (Table~\ref{tab:res}). The measured scatters on the VST data sets (both original and re-calibrated) are consistent with the reported scatter against SDSS: 0.014 mags for KiDS \citep{kids_data} and 0.013 mags for ATLAS \citep{atlas_data}. The NSC \textit{g}-band is reported to have a zero-point scatter (relative to its mean zero-point) of $\sim 0.039$ mags based on $\sim 9000$ fields \citep{nsc}. We find it to be $\sim 0.007$ w.r.t. SDSS over the 72 tiles used in this work, which after the re-calibration remains at the same level. In sum, the re-calibration with Refcat2 improves the \textit{g}-band zero-point accuracy, and keeps the same level of precision (comparing with SDSS), or improves it (comparing with PanSTARRS).

\subsection{The \textit{u}-band calibration}
\label{subsec:calib}

The KiDS {\it u}-band calibration initially uses the stellar locus \citep[namely, the s-color method, ][]{s_color, s_color_cali} to determine the $(u-r)$ zeropoint of each field \citep{kids_450}, and calibrates {\it r}-band from Gaia DR2 \citep{kids}. However, they find a residual latitude dependence compared to SDSS, which we qualitatively reproduce in App.~\ref{sec:appe}.  Independent of our work (which had started prior to the KiDS DR4 release), they implemented a latitude-dependent correction based on the measured colors of blue stars.  Despite the parallels, there are important differences between the KiDS {\it u}-band calibration and our blue-edge method: (1) our method uses a specific feature of the color distribution of stars, which can be found even on uncalibrated data; whereas KiDS utilizes all stars within a specific $(g-r)$ range.  (2)  The functional form of the latitude dependence is different.  (3) We calibrate our method on the entirety of SDSS, rather than just the KiDS-SDSS overlap.  \citet{kids} assign a residual \textit{u}-band uncertainty of up to 0.05 mags, which we believe we have significantly improved upon.

The ATLAS DR4 {\it u}-band photometry is calibrated from the subset of fields taken under photometric conditions by solving for the zero-point of each field based on stars overlapping on the field edges.  The zero-point of the anchor fields is the nightly photometric calibration based on standard star field observations \citep{atlas_web}.

The NSC {\it u}-band photometry is calibrated from a color equation which predicts DECam {\it u-}band from GALEX NUV \citep{galex}, Gaia DR1 {\it G-}band and \citep{gaia_dr1} 2MASS {\it J}-band \citep{2mass}, based on overlapping data in the SDSS Stripe 82 area \citep{nsc}.

To (re-)calibrate {\it u}-band data with the blue-edge method, we assume that the contamination factor $D$ in Eq.~\ref{equ:fit} is the same for all instruments of SDSS-like filters, and keep it fixed at 0.04 from the SDSS joint fit (Table~\ref{tab:sdss}). The halo color $x_0^{\mathrm{halo}}$ may vary for different facilities, and needs to be determined in order to use equation~(\ref{equ:fit}) for calibrating \textit{u}-band zero-points based on observed edge colors. 
Notice that $x_0^{\mathrm{halo}}$ cannot be determined a priori (e.g. via a color transform from SDSS to other instruments), since  $x_0^{\mathrm{halo}}$ is not the $u-g$ color of a specific star, but the edge color of a population, therefore it is difficult to define the corresponding colors (e.g. $g-i$) involved in the \textit{ug}-band color equations (equation ~\ref{eqn:color_synth}).
However, $x_0^{\mathrm{halo}}$ is straightforward to establish empirically by comparing with SDSS fields (which we take as truth values). Namely, because the {\it u}-band color transformation $(u_{\rm inst} - u_{\rm SDSS})$ between the instrument in question and SDSS depends only on the SDSS $(g-i)$ color, any residual {\it u}-band offset with SDSS can simply be negated by adjusting $x_0^{\mathrm{halo}}$.
In practice, we therefore simultaneously determine $x_0^{\mathrm{halo}}$ and calibrate the \textit{u}-band photometry of the dataset in question. Technically, this procedure works on a single field overlapping SDSS, but we suggest performing it with as many tiles as possible to mitigate possible bias rooted in scatters of SDSS photometry. 

Since KiDS and ATLAS are both taken with VST, we use both to determine the relation for VST:
\begin{enumerate}
    \item For a given field, calibrate the KiDS/ATLAS \textit{g}-band magnitudes with Refcat2 as in Sect.~\ref{subsec:gdis}.
    \item Detect the $(u-g)$ edge colors with the calibrated \textit{g}-band, as described in Sect.~\ref{subsec:iso}. Notice that these colors are uncalibrated, as the {\it u}-band is still uncalibrated.
    \item For each field, calculate the residual between the measured $(u-g)$ edge color and the model prediction equation~(\ref{equ:fit}), setting the contamination factor $D$ to 0.04, and the halo color $x_0^{\mathrm{halo}}$ to zero for an initial pass.
    \item Calibrate the \textit{u}-band magnitudes from the residuals in (iii) as the zero-points for each field. (Recall that the method calibrates the {\it u}-band zero-point by ``moving'' each field onto the best-fit relation, equation~[\ref{equ:fit}]). 
    \item Compare the calibrated \textit{u}-band magnitudes with SDSS using the synthetic color transform (\ref{eqn:color_synth}), and calculate the residual median offsets per field. The halo color $x_0^{\mathrm{halo}}$ is the mean of the distribution of median offsets.
\end{enumerate}
This procedure yields the calibrated color-latitude relation for VST:
\begin{equation}
	\quad\quad\quad\quad\quad  \left( u - g \right)_{_{\mathrm{VST}}} = \frac{0.04}{|\sin{b}|} + 0.697.
	\label{eqn:calib_vst}
\end{equation}
The procedure for the NSC catalog is exactly the same. The resulting color-latitude equation for DECam is:
\begin{equation}
	\quad\quad\quad\quad\quad  \left( u - g \right)_{_{\mathrm{DECam}}} = \frac{0.04}{|\sin{b}|} + 0.504.
	\label{eqn:calib_nsc}
\end{equation}
With the above instrument-specific curves, new datasets taken with VST or DECam can be straightforwardly calibrated:
\begin{enumerate}
    \item Calibrate the \textit{g}-band onto the SDSS AB system (\ref{subsec:gdis}).
    \item Detect the halo MSTO edge color as described in Sect.~\ref{subsec:iso}. 
    \item Compare with equation (\ref{eqn:calib_vst}) or (\ref{eqn:calib_nsc}) depending on instrument. Calibrate onto the curve.
\end{enumerate}

The resulting \textit{u}-band offset distributions are shown in the right column of Fig.~\ref{fig:res}. Similar to the \textit{g}-band panels, the offset distributions from the original data sets are shown in orange, while the re-calibrated histograms in blue. The orange/blue histograms are offsets derived with the synthetic color equations (\ref{eqn:color_synth}), while the empty histograms are the original data sets with color equations derived from overlapping fields (equation [\ref{eqn:color}]) for comparison. The standard deviation of the empty and the blue distributions are shown in corresponding entries in Table~\ref{tab:res}.

The original KiDS/ATLAS {\it u}-band zeropoint precision is reported for both to be $\sim 0.035$~mags RMS scatter relative to SDSS \citep[see the release notes][]{kids_data, atlas_data}. We first produce our own estimate of this value from the KiDS/ATLAS data included in this work, transforming the original magnitude to SDSS using the color transformations that we fit to overlapping data (equation [\ref{eqn:color}]), i.e. the empty histograms in Fig.~\ref{fig:res}.  We find a standard deviation of 0.033 mags for KiDS and 0.027 mags for ATLAS. The smaller scatter for ATLAS is most likely due to the larger tile size used in this work. Similarly, a global NSC {\it u}-band scatter of 0.07 mags is reported in \citet{nsc} while we find it to be 0.041 mags in the regions overlapping SDSS. 

After the re-calibration, the standard deviation is reduced to $\sim 0.016$ mags for KiDS, 0.020 mags for ATLAS and 0.016 mags for NSC (these are based on the synthetic color equations [\ref{eqn:color_synth}]), i.e. the {\it u}-band precision is substantially improved for all three datasets.

\section{Summary and Discussion}
\label{sec:dis}

In this paper we have presented a method for the photometric calibration of \textit{u}-band imaging based on the “blue tip" of Galactic halo stars. The method requires no standard star comparison, nor does it require the fields to overlap with each other. This makes our method powerful even for pointed, single-field observations. It is applicable to fields outside the galactic plane ($|b|>30^{\circ}$), for which both {\it u}-band and {\it g}-band imaging are available, deep enough to detect the halo MSTO population ($g \approx 21$).
Based on comparing three validation datasets (KiDS, ATLAS, and NSC) with SDSS {\it u}-band photometry, we reach a competitive precision for the zero-point calibration ($\le 0.02$ mag), satisfying the Rubin-LSST photometry requirement.

Our method is based empirically on the slow spatial variation of \textit{u-g} color of halo MSTO stars. We develop an algorithm to detect the edge \textit{u-g} color of halo MSTOs (Sect.~\ref{subsec:iso}), and observe that this color is strongly correlated with galactic latitudes. We fit a simple plane-parallel model to the correlation (Sect.~\ref{subsec:con}), and use it for the \textit{u}-band calibration. The model is motivated by contamination of disk stars in the halo MSTO sample.  It is worth noting that the photometric calibration of the Javalambre Photometric Local Universe Survey DR2 \citep[J-PLUS;][]{jplus, jplus_cali} shares a similar flavor to our method, despite of a different approach and motivation. J-PLUS DR2 first builds a median-metallicity stellar locus from a spectroscopic sample, then model the deviance of observed color to the locus as a function of milky way locations. A precision of 18 mmag in \textit{u}-band at a magnitude limit of 21.16 is reported for the final calibration.  The deviance-latitude dependence found by J-PLUS \citep[Fig.~6 of][]{jplus_cali} is notably similar to the color-latitude relation that we utilize here, which is attributed to stellar metallicity variation rather than contamination of a halo stellar population by disk stars.  We emphasize that the color-latitude relation we find is entirely empirical, and that the interpretation of the underlying physical cause of the color-latitude relation, equation (\ref{equ:fit}), has no effect on our method itself, other than the assumption that the relation holds in both hemispheres. 

Sect.~\ref{subsec:calib} details how to establish the color-latitude relation for a specific instrument. Once the relation is found, it can be used to calibrate any data taken with the instrument. The magnitude transform to SDSS is also needed in the procedure, and can be derived with synthetic magnitudes as described in App.~\ref{sec:appc} given the filter responses.  The method can then be through the following steps:
\begin{enumerate}
    \item For a given field, calibrate the \textit{g}-band with ATLAS-Refcat2 as described in Sect.~\ref{subsec:gdis}.
    \item Detect the halo MSTO ``blue tip'' \textit{u-g} color using model (\ref{eqn:conv}).
    \item Calibrate the detected color onto the color-latitude relation (equation [\ref{eqn:calib_vst}] for Omegacam and equation [\ref{eqn:calib_nsc}] for DECam). In other words, use the residual between the detected \textit{u-g} color and the color-latitude relation as the \textit{u}-band zero-point.
\end{enumerate}
In this work, we present calibrated color-latitude relations for Omegacam and DECam, but the method can be straightforwardly expanded to other instruments (provided that some fields overlapping SDSS are available in order to determine $x_0^{\rm halo}$) following the procedure discussed in Sect.~\ref{subsec:calib}.

We have verified the method by re-calibrating the \textit{u}-band photometry of the KiDS, ATLAS and NSC datasets, and find an improved zeropoint precision (when comparing to SDSS as the truth dataset) for all of them (Sect.~\ref{subsec:calib}).  For KiDS/ATLAS/NSC, we find RMS scatters of 0.016/0.020/0.016 mag against SDSS, compared to the RMS scatters with the original calibration of 0.033/0.027/0.041 mag.

As the method includes (re-)calibrating \textit{g}-band imaging of the same field, using ATLAS-Refcat2 as the \textit{g}-band reference, we also examine the re-calibrated vs. original \textit{g}-band photometry of KiDS/ATLAS/NSC. We show that for all three, the re-calibration is in excellent agreement with PanSTARRS (as expected, given that ATLAS-Refcat2 relies heavily on PanSTARRS), and remains at the same precision level with respect to SDSS as the original \textit{g}-band calibration.

We also perform a few additional validation of our method detailed in App.~\ref{sec:appd} and App.~\ref{sec:appe}. In App.~\ref{sec:appd}, we attempt to reject disk stars from the sample of halo MSTO stars using Gaia parallax and proper motion measurements. We find that Gaia EDR3 is not yet deep enough to substantially reduce the disk star contamination in our sample. In App.~\ref{sec:appe} we briefly compare our calibration with two stellar locus based methods. We find a similar latitude dependence from both methods based on an SDSS self-calibration set up.

One of the applications that could benefit from this work is photometric redshift measurements of large scale cosmological imaging surveys. For example, the KiDS-1000 cosmology is based on a nine-band photometric redshift estimate \citep[\textit{ugriZYJH$K_s$};][]{kids_1000_res, kids_1000_pz}, where the \textit{u}-band has the largest zero-point scatter out of the 4 optical bands (0.035, 0.014, 0.005, and 0.009 mags for \textit{ugri} respectively; see the DR4 release note \citet{kids_data}). Our re-calibration method reduces the \textit{u}-band scatter by about factor of two, which should improve the photometric redshift estimation.

This work has potential benefits for the Rubin Observatory's LSST. The current photometric calibration plan is to adopt a similar pipeline as the Forward Global Calibration Method (FGCM) by \citet{fgcm}. FGCM combines auxiliary data with survey images and models the atmosphere and system throughput of all bands at any survey time and location. FGCM has delivered a stable photometric calibration of $6-7$ mmags random error per exposure for the first three years of the Dark Energy Survey \citep[DES;][]{des} in the \textit{grizY} bands. However, since DES does not include \textit{u}-band observations, the performance of FGCM on \textit{u}-band, which depends sensitively on the near-UV cut-off of atmospheric transparency, is to this date unknown. Our method provides a fast calibration of the LSST \textit{u}-band imaging which can be used as a preliminary on-the-fly calibration, an independent cross-check of FGCM, and/or be incorporated into the FGCM scheme. Moreover, the large Southern sky footprint of LSST provides complementary coverage to SDSS in the North, which will fact-check our assumption of a whole-sky color-latitude relation. Furthermore, as LSST is deeper than SDSS, it will allow us to study the magnitude dependence of the MSTO blue-tip color, and thus the stellar structure in the Milky Way halo.
\section*{Acknowledgements}
The authors thank the anonymous referee for their constructive comments on the paper.  We also thank
Ricardo Herbonnet, Robyn E. Sanderson, Robert Lupton, and Eli Rykoff for illuminating discussion and suggestions. 

SL and AvdL are supported by the U.S. Department of Energy under award DE-SC0018053. This work is also supported by the U.S. Department of Energy, Office of High Energy Physics under grant number DE-1161130-116-SDDTA.

This research made use of Astropy, a community-developed core python package for Astronomy \citep{astropy}, as well as TOPCAT \citep{topcat}, a graphical user tool for interactive table manipulation, and STILTS \citep{stilts}, a command-line based tool for the processing of tabular data. Other packages used in this work include CasJobs \citep{casjobs}, Numpy \citep{np}, Matplotlib \citep{plt}, and pystan \citep{pystan}.

Funding for the SDSS and SDSS-II has been provided by the Alfred P. Sloan Foundation, the Participating Institutions, the National Science Foundation, the U.S. Department of Energy, the National Aeronautics and Space Administration, the Japanese Monbukagakusho, the Max Planck Society, and the Higher Education Funding Council for England. The SDSS Web Site is \url{http://www.sdss.org/}.
The SDSS is managed by the Astrophysical Research Consortium for the Participating Institutions. The Participating Institutions are the American Museum of Natural History, Astrophysical Institute Potsdam, University of Basel, University of Cambridge, Case Western Reserve University, University of Chicago, Drexel University, Fermilab, the Institute for Advanced Study, the Japan Participation Group, Johns Hopkins University, the Joint Institute for Nuclear Astrophysics, the Kavli Institute for Particle Astrophysics and Cosmology, the Korean Scientist Group, the Chinese Academy of Sciences (LAMOST), Los Alamos National Laboratory, the Max-Planck-Institute for Astronomy (MPIA), the Max-Planck-Institute for Astrophysics (MPA), New Mexico State University, Ohio State University, University of Pittsburgh, University of Portsmouth, Princeton University, the United States Naval Observatory, and the University of Washington.

Based on data products from observations made with ESO Telescopes at the La Silla Paranal Observatory under programme IDs 177.A-3016, 177.A-3017 and 177.A-3018, and on data products produced by Target/OmegaCEN, INAF-OACN, INAF-OAPD and the KiDS production team, on behalf of the KiDS consortium. OmegaCEN and the KiDS production team acknowledge support by NOVA and NWO-M grants. Members of INAF-OAPD and INAF-OACN also acknowledge the support from the Department of Physics and Astronomy of the University of Padova, and of the Department of Physics of Univ. Federico II (Naples)

The Pan-STARRS1 Surveys (PS1) and the PS1 public science archive have been made possible through contributions by the Institute for Astronomy, the University of Hawaii, the Pan-STARRS Project Office, the Max-Planck Society and its participating institutes, the Max Planck Institute for Astronomy, Heidelberg and the Max Planck Institute for Extraterrestrial Physics, Garching, The Johns Hopkins University, Durham University, the University of Edinburgh, the Queen's University Belfast, the Harvard-Smithsonian Center for Astrophysics, the Las Cumbres Observatory Global Telescope Network Incorporated, the National Central University of Taiwan, the Space Telescope Science Institute, the National Aeronautics and Space Administration under Grant No. NNX08AR22G issued through the Planetary Science Division of the NASA Science Mission Directorate, the National Science Foundation Grant No. AST-1238877, the University of Maryland, Eotvos Lorand University (ELTE), the Los Alamos National Laboratory, and the Gordon and Betty Moore Foundation.

This work has made use of data from the European Space Agency (ESA) mission {\it Gaia} (\url{https://www.cosmos.esa.int/gaia}), processed by the {\it Gaia} Data Processing and Analysis Consortium (DPAC, \url{https://www.cosmos.esa.int/web/gaia/dpac/consortium}). Funding for the DPAC has been provided by national institutions, in particular the institutions participating in the {\it Gaia} Multilateral Agreement.
\section*{Data Availability}

The data sets underlying this article are publicly available. They can be accessed at:
\begin{enumerate}
    \item[] SDSS: from the SkyServer at \url{http://skyserver.sdss.org/dr15/en/tools/search/sql.aspx}
    \item[] KiDS: from the KiDS Data Release 4 website \url{https://kids.strw.leidenuniv.nl/DR4/access.php}
    \item[] ATLAS: from the OmegaCAM Science Archive (OSA) website \url{http://osa.roe.ac.uk/}
    \item[] NSC: through the TAP query at \url{https://datalab.noirlab.edu/docs/manual/UsingTheNOAODataLab/DataAccessInterfaces/CatalogDataAccessTAPSCS/CatalogDataAccessTAPSCS.html}
    \item[] PanSTARRS: through CasJobs at \url{https://mastweb.stsci.edu/ps1casjobs/home.aspx}
    \item[] ATLAS-Refcat2: through CasJobs at \url{https://archive.stsci.edu/prepds/atlas-refcat2/}
    \item[] SHELA: as the supplementary data of \citet{shela} at \url{https://iopscience.iop.org/article/10.3847/1538-4365/aaee85/data}
\end{enumerate}

For more details about data acquirement, please see APP.~\ref{sec:appa}.




\bibliographystyle{mnras}
\bibliography{ref.bib} 



\appendix
\section{Detailed Data Acquirement}
\label{sec:appa}

In this section, we provide detailed data queries and selections for the data sets used in this work, as briefly introduced in Sect.~\ref{sec:data}. We also show the selections for SHELA data \citep{shela}, a DECam data set that is included in NSC but with an independent calibration. SHELA is not used in any of the calibration work, but only to compare its photometry with NSC.

\subsection{SDSS data}
\label{subsec:sdss_data}
We download SDSS data from the SkyServer\footnote{\url{http://skyserver.sdss.org/dr15/en/tools/search/sql.aspx}} via a python package urllib. SDSS photometry is given in luptitudes\footnote{\url{https://www.sdss.org/dr15/algorithms/magnitudes/\#asinh}}, as
introduced by \citet{luptitude}. To be consistent with the validation set, we take the SDSS PSF flux density and derive the (Pogson) AB
magnitudes \citep{ab_mag} and the errors. Let {\tt X} stand for all of the \textit{u, g, r, i} and \textit{z} bands. we calculate
$$
\begin{aligned}
	{\tt X}_{_{\mathrm{SDSS}}} &  = 22.5 - 2.5 \log_{10}({\tt psfFlux\_X}) - {\tt extinction\_X} \\
	{\tt X}\_\mathrm{err}_{_{\mathrm{SDSS}}} & = 1.085 / sqrt({\tt psfFluxIvar\_X}) / {\tt psfFlux\_X}.
\end{aligned}
$$
where {\tt X}\_{SDSS}, {\tt X}\_err\_{SDSS} are the {\tt X}-band magnitude and the error,
{\tt psfFlux\_X} and {\tt psfFluxIvar\_X} are the {\tt X}-band PSF flux density and the inverse variance, and {\tt extinction\_X}
is the {\tt X}-band extinction. The flux density is measured in the units of nanomaggies\footnote{\url{https://www.sdss.org/dr15/help/glossary/\#N}}
which corresponds to a zero-point of 22.5 in magnitude. In addition, the SDSS $u$-band magnitude differs from a perfect AB magnitude by
0.04\footnote{\url{https://www.sdss.org/dr13/algorithms/fluxcal/\#SDSStoAB}}, so we subtract 0.04 from the magnitude. Similarly, we add 0.02 mags to the {\it z}-band. We use only the clean photometry
identified from SDSS by setting the flag
$$ 
{\tt stars.clean}=1,
$$
and we keep only the sources with magnitude of
$$
14 < {\tt X}_{\mathrm{SDSS}} < 21,
$$
and the magnitude error < 0.1 in all bands, 
$$
{\tt X}\_\mathrm{err}_{_{\mathrm{SDSS}}} < 0.1.
$$
An example query is as follows:
\begin{verbatim}
select ra,dec,l,b,\
-2.5*LOG10(psfFlux_u) + 22.5 
- extinction_u - 0.04 as psfPogCorr_u,
-2.5*LOG10(psfFlux_g) + 22.5 \
- extinction_g + 0.00 as psfPogCorr_g, 
-2.5*LOG10(psfFlux_r) + 22.5 \
- extinction_r + 0.00 as psfPogCorr_r, 
-2.5*LOG10(psfFlux_i) + 22.5 \
- extinction_i + 0.00 as psfPogCorr_i, 
-2.5*LOG10(psfFlux_z) + 22.5 \
- extinction_z + 0.02 as psfPogCorr_z, 
1.085/SQRT(psfFluxIvar_u)/psfFlux_u as psfPogErr_u, 
1.085/SQRT(psfFluxIvar_g)/psfFlux_g as psfPogErr_g, 
1.085/SQRT(psfFluxIvar_r)/psfFlux_r as psfPogErr_r, 
1.085/SQRT(psfFluxIvar_i)/psfFlux_i as psfPogErr_i, 
1.085/SQRT(psfFluxIvar_z)/psfFlux_z as psfPogErr_z  
from star as s where ra between 0 
and 1 and dec between 0 and 1
and s.clean=1
and -2.5*LOG10(psfFlux_u) + 22.5 
- extinction_u - 0.04 between 14 and 21
and -2.5*LOG10(psfFlux_g) + 22.5 \
- extinction_g  between 14 and 21
and -2.5*LOG10(psfFlux_r) + 22.5 \
- extinction_r  between 14 and 21
and -2.5*LOG10(psfFlux_i) + 22.5 \
- extinction_i  between 14 and 21
and -2.5*LOG10(psfFlux_z) + 22.5 \
- extinction_z + 0.02 between 14 and 21
and 1.085/SQRT(psfFluxIvar_u)/psfFlux_u < 0.1 
and 1.085/SQRT(psfFluxIvar_g)/psfFlux_g < 0.1 
and 1.085/SQRT(psfFluxIvar_r)/psfFlux_r < 0.1 
and 1.085/SQRT(psfFluxIvar_i)/psfFlux_i < 0.1
and 1.085/SQRT(psfFluxIvar_z)/psfFlux_z < 0.1
\end{verbatim}

\begin{figure}
	\includegraphics[width=\columnwidth]{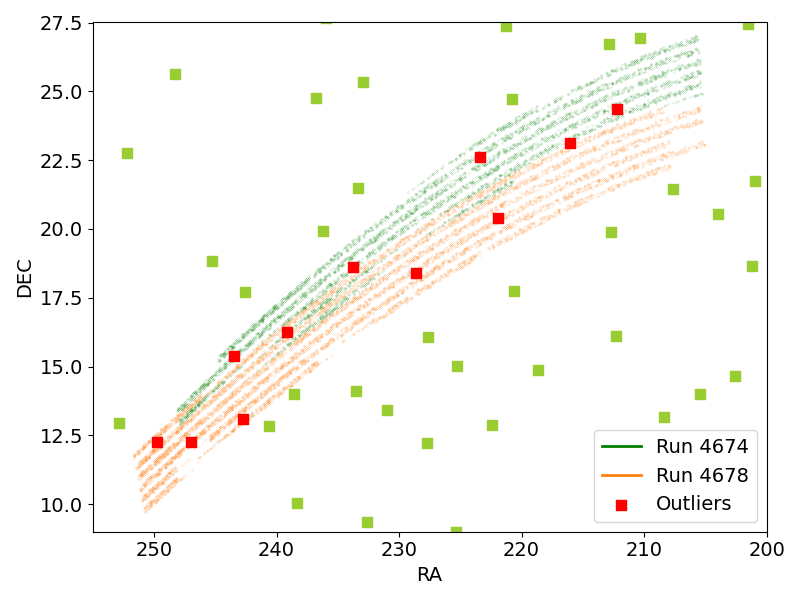}
	\caption{Sky footprint of some SDSS data used in this work. Each square is a $2\times2^\circ$ tile. The red tiles are outliers in the color-latitude relation (Fig.~\ref{fig:pos}). They are excluded from the fitting of model (\ref{equ:fit}).}
	\label{fig:runs}
\end{figure}

Notice that some outliers are found in the left panel of Fig.~\ref{fig:pos}. Most of them are exposures taken in two specific SDSS runs (4674 and 4678, see Fig.~\ref{fig:runs}). They are excluded from the MCMC fitting.

\subsection{KiDS data}
Kids data are batch-downloaded with the bash script provided on the KiDS Data Release 4 website\footnote{\url{https://kids.strw.leidenuniv.nl/DR4/access.php}}.
We use the KiDS Multi-band catalog (DR4.1 update), and select stars in each tile with high-fidelity photometry by the following criteria: ({\tt X} stands for all of the \textit{ugri}-bands:
\begin{enumerate}
\item {\tt Flag} = 0 \\
	{\tt Flag} is the $r$-band photometric flag from SExtractor\footnote{\url{https://www.astromatic.net/software/sextractor}}. The value 0 means the source detected has good photometric property (not over-exposed, truncated nor blended with nearby sources).
\item {\tt SG2DPHOT} = 5 \\
	This is the star-galaxy classification based on $r$-band photometry. The value 5 means the source is a high confidence star candidate.
\item {\tt FLAG\_GAaP\_X} = 0 \\
	The GAaP photometry flag. It is set to 100 when no GAaP flux is measured, otherwise 0 \citep{gaap}.
\item 14 < {\tt MAG\_GAaP\_X} < 21 \\
    We keep sources with magnitude within the range, and
\item {\tt MAGERR\_GAaP\_X} < 0.1 \\
	magnitude errors smaller than 0.1.
\end{enumerate}

\subsection{ATLAS data}
We obtain the ATLAS data via the Freeform Query on the OmegaCAM Science Archive (OSA) website\footnote{\url{http://osa.roe.ac.uk/}}.
We apply the following selection criteria per star on ATLAS data ({\tt X} stands for all of the \textit{ugriz}-bands):
\begin{enumerate}
	\item {\tt XClass} = -1 \\
		{\tt XClass} is the most probable morphological classification in each band with -1=stellar, +1=non-stellar, 0=noise and -2=borderline stellar.
	\item 14 < {\tt XAperMag3} < 21 \\
		ATLAS has many bright stars that are not needed for our purpose so we exclude them. Note that ATLAS doesn't offer PSF magnitudes. We use {\tt XAperMag3} as suggested by \citet{atlas}.
	\item {\tt XAperMag3Err} < 0.1 \\
		We keep only sources with small magnitude errors.
\end{enumerate}
The the extinction is corrected with 
$$
	{\tt MAG\_X} = {\tt XAperMag3} - {\tt aX}.
$$
An example query is shown below:
\begin{verbatim}
select ra, dec, l, b,
uAperMag3 - aU as u, gAperMag3 - aG as g,
rAperMag3 - aR as r, iAperMag3 - aI as i,
zAperMag3 - aZ as z,
uAperMag3Err as uerr, gAperMag3Err as gerr,
rAperMag3Err as rerr, iAperMag3Err as ierr,
zAperMag3Err as zerr,
from atlasSource where ra between 0 and 62 
and uClass=-1 and gClass=-1 and rClass=-1
and iClass=-1 and zClass=-1
and uAperMag3Err < 0.1 and gAperMag3Err < 0.1
and rAperMag3Err < 0.1 and iAperMag3Err < 0.1
and zAperMag3Err < 0.1
and uAperMag3 between 14 and 21
and gAperMag3 between 14 and 21
and rAperMag3 between 14 and 21
and iAperMag3 between 14 and 21
and zAperMag3 between 14 and 21
\end{verbatim}

\subsection{NSC data}
Since the NSC catalog has limited sources, we use only the \textit{ug}-bands selection to maximize the amout of data.
We apply the following cuts:
\begin{enumerate}
	\item 14 < ({\tt umag, gmag}) < 21 \\
		We keep sources with both magnitudes within the range.
	\item ({\tt uerr, gerr}) < 0.1 \\
		The uncertainty cut on both magnitudes.
	\item {\tt flags = 0} \\
	    The photometry flags.
	\item $1\arcsec<${\tt fwhm}$<1.6\arcsec$ \\
	    We select stars with the object size, and
	\item {\tt class\_star} > 0.95 \\
	    the star classifier by SExtractor.
\end{enumerate}
We query all stars from \texttt{nsc\_dr2.object} using STILTS \citep{stilts}, a tool set for STIL (the Starlink Tables Infrastructure
Library) via the Table Access Protocol
(TAP\footnote{\url{https://www.ivoa.net/documents/TAP/}}). Stars are binned according to exposure centers found in
\texttt{nsc\_dr2.exposure}. The bash command that queries all NSC stars for this work is shown below:
\begin{verbatim}
stilts tapquery \ 
tapurl='http://datalab.noao.edu/tap' \
adql="SELECT id, ra, dec, glon, glat, \
umag, urms, uerr, gmag, grms, gerr, \
fwhm, class_star, ebv \
FROM nsc_dr2.object where (glat > 30 or glat < -30) 
and dec > -30 and flags=0 \
and umag > 14 and umag < 21 and uerr < 0.1 \
and gmag > 14 and gmag < 21 and gerr < 0.1 \
and fwhm>1 and fwhm<1.6 and class_star>0.95" \
out=raw.fits
\end{verbatim}
and the bash command for querying the exposure centers is below:
\begin{verbatim}
stilts tapquery \
tapurl='http://datalab.noao.edu/tap' \
adql="SELECT ra, dec, glon, glat, mjd, \
exposure, exptime, airmass, nmeas, \ 
zptype, zpterm, depth10sig \
FROM nsc_dr2.exposure where filter='u'  and \
(glat > 27 or glat < -27) and dec > -29.7" \
out=center.fits
\end{verbatim}

We correct the reddening with extinction coefficients found in \citet{decam_ext}. 

\subsection{PanSTARRS data}
We use CasJobs to query all PanSTARRS objects that cover as much as possible the KiDS, ATLAS and NSC footprint from table {\tt MeanObject}, and cross-match with table {\tt ObjectThin} for the selection cuts. The cuts are ({\tt X} stands for all of the {\it g, r, i} and {\it z} bands):
\begin{enumerate}
	\item 14 < {\tt XMeanPSFMag} < 21 \\
		For stars, we use {\tt MeanPSFMag}, the mean PSF magnitude that has the lowest noise \citep{pan_data}\footnote{\url{https://outerspace.stsci.edu/display/PANSTARRS/PS1+FAQ+-+Frequently+asked+questions\#PS1FAQ-Frequentlyaskedquestions-WhichmagnitudesshouldIuse?}}.
		We keep stars in this range for all bands.
	\item $-0.2 < {\tt iMeanPSFMag - iMeanKronMag} < 0.05$ \\
		We keep objects whose {\it i}-band PSF magnitude and Kron Magnitude show moderate differences. This cut serves as an efficient star-galaxy
		separation \citep{pan_sep}\footnote{\url{https://outerspace.stsci.edu/display/PANSTARRS/How+to+separate+stars+and+galaxies}}.
	\item {\tt XMeanPSFMagErr} < 0.1 \\
		We keep sources with small magnitude errors.
	\item {\tt nX} > 0 \\
		{\tt nX} is the number of single epoch detections in filter {\tt X}. We require the object to be detected at least once
		in each filter.
\end{enumerate}
An example query is shown below: 
\begin{verbatim}
select o.raMean as ra, o.decMean as dec, 
m.gMeanPSFMag, m.rMeanPSFMag, m.iMeanPSFMag,
m.zMeanPSFMag, m.gMeanPSFMagErr, m.rMeanPSFMagErr, 
m.iMeanPSFMagErr, m.zMeanPSFMagErr 
into my_database 
from fGetObjFromRectEq(322, -30, 360, -9 ) nb 
inner join ObjectThin o 
on o.objid=nb.objid and o.nDetections>1 
and ng > 0 and nr > 0 and ni > 0 and nz > 0 
inner join MeanObject m on o.objid=m.objid 
and m.gMeanPSFMag > 14 and m.rMeanPSFMag > 14 
and m.iMeanPSFMag > 14 and m.zMeanPSFMag > 14 
and m.gMeanPSFMag < 21 and m.rMeanPSFMag < 21 
and m.iMeanPSFMag < 21 and m.zMeanPSFMag < 21
and m.gMeanPSFMagErr < 0.1 and m.rMeanPSFMagErr < 0.1 
and m.iMeanPSFMagErr < 0.1 and m.zMeanPSFMagErr < 0.1 
and iMeanPSFMag - iMeanKronMag > -0.2 
and iMeanPSFMag - iMeanKronMag < 0.05
\end{verbatim}

We correct the extinction with band coefficients found in \citet{3d_dust}.

\subsection{ATLAS-Refcat2 data}
We use CasJobs to query from the table {\tt refcat}, with the same general cuts on the magnitudes ({\tt X}={\it griz})
$$
    14 < {\tt X} < 21, \quad \mathrm{and} \quad {\tt dX} < 0.1
$$
where {\tt dX} is the magnitude uncertainty. We use the SFD extinction map with PanSTARRS band coefficients to correct for reddening.
A simple example query:
\begin{verbatim}
select RA, DEC, g, r, i, z, Ag,
dg, dr, di, dz into my_database from refcat2	
where RA between 150 and 233.5 
and	DEC between -20 and -10 
and g > 14 and g < 21 and r > 14 and r < 21 
and i > 14 and i < 21 and z > 14 and z < 21 
and	dg < 0.1 and dr < 0.1 and di < 0.1 and dz < 0.1
\end{verbatim}

\subsection{SHELA data}
The entire DECam-SHELA catalog is provided as supplementary data for \citet{shela}\footnote{\url{https://iopscience.iop.org/article/10.3847/1538-4365/aaee85/data}}. We keep sources with the internal and external flags (SExtractor-based) of 0 in all bands (\textit{ugriz}), and calculate the magnitudes and uncertainties from the flux measurements. The fluxes are calibrated and provided in the physical units of $\mu$Jy, corresponding to a magnitude zero-point of 23.9. The stars are selected by the FWHM size: {\tt r0.5 < 1} arcsec. We again keep sources with magnitude 14 < {\tt X} < 21 and uncertainties {\tt dX} < 0.1, with {\tt X} standing for all \textit{ugriz}-bands.

\section{Color Equations from Photometry}
\label{sec:appb}

For completeness, we fit the color equations between the validation sets (KiDS/ATLAS/SDSS) and the reference catalogs (SDSS/PanSTARRS/Refcat2) on the cross-matched samples, and show that some inconsistencies can be found between the data sets. Notice that the fitted color equations in this section are not used in the calibration procedure due to the inconsistencies, though they are used for simple comparison purposes to form the ``empty'' histograms in Fig.~\ref{fig:res}.

\begin{figure*}
        \includegraphics[width=0.65\columnwidth]{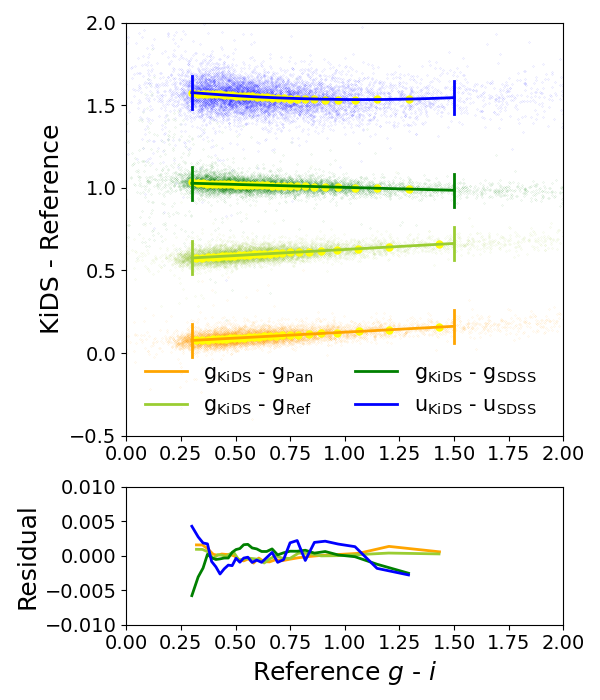}
        \includegraphics[width=0.65\columnwidth]{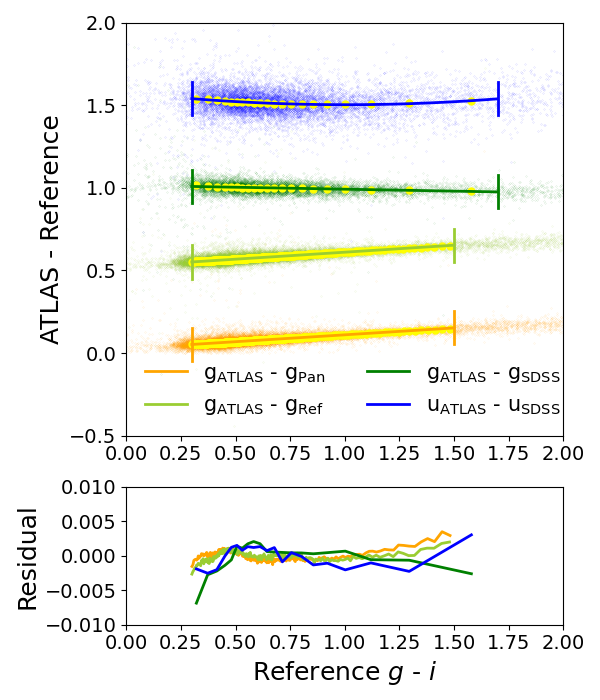}
        \includegraphics[width=0.65\columnwidth]{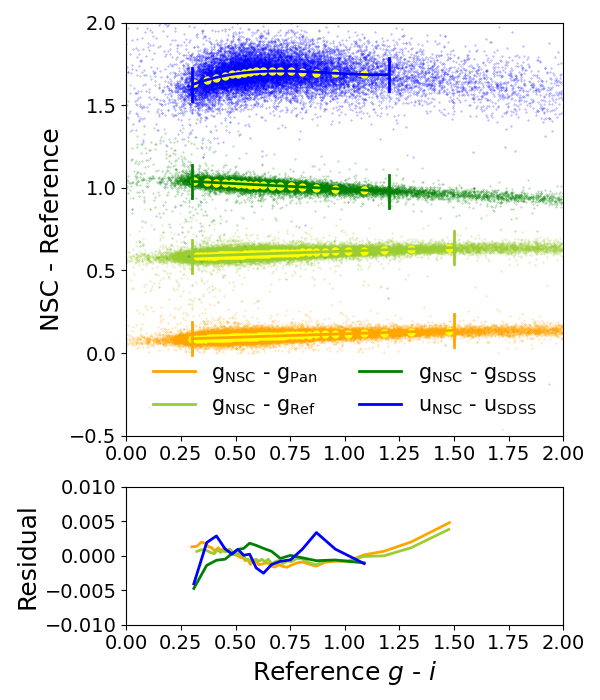}
	\caption{Estimating the color transformation equation on the original photometry of KiDS/ATLAS/NSC. Data points are plotted with offsets (each set is 0.5 higher than the one below it) for illustration. Notice that these color equations are not used in the calibration procedure. Yellow dots are the median magnitude differences (with uncertainties) in each bin, obtained from bootstrap resampling. The color equations are fit to the yellow dots. A 2nd order polynomial is used for the {\it u}-band transform for KiDS/ATLAS, a 3rd order for DECam {\it u}-band, and 1st order for all the {\it g}-bands. The color transforms are given in equation (\ref{eqn:color}), and the fitting ranges are listed in Table~\ref{tab:range}.}
	\label{fig:color}
\end{figure*}

We first bin all the matched stars in the reference \textit{g-i} color (SDSS, Refcat2 or PanSTARRS). For the SDSS-matched samples, this covers: The entire KiDS North, part of ATLAS (both in the North and the South, see the yellow region in Fig.~\ref{fig:cover}), and tiles of NSC that fall in the SDSS region. For the PanSTARRS-matched samples, this consists of all the data above D.E.C. $>-30^\circ$: KiDS North and half of KiDS South, most of ATLAS and all of NSC (the NSC sample is selected to have D.E.C. $>-30^\circ$). For Refcat2, this includes all the data sets. We bin the matched samples with an adaptive bin width to include 10000 stars in each bin. We then bootstrap the stars to estimate the median magnitude difference with uncertainty in each bin, to which we fit polynomials as a function of the median reference \textit{g-i} color of each bin. The best-fit polynomials are shown in Fig.~\ref{fig:color}. 

We use 2nd order color terms for the SDSS \textit{u}-band, and a linear regression for all the \textit{g}-bands. For the DECam \textit{u}-band we use a 3rd order polynomial. In each case, the order of the polynomial is chosen to best describe the bulk shape of the stars in Fig.~\ref{fig:color}.  We find the following color terms:
\begin{equation}
	\begin{aligned}
		u_{\mathrm{KiDS}} - u_{\mathrm{SDSS}} & =  0.069 x_S^2 - 0.149x_S + 0.114, \\
		u_{\mathrm{ATLAS}} - u_{\mathrm{SDSS}} & = 0.074 x_S^2 - 0.148 x_S + 0.077, \\
		u_{\mathrm{NSC}} - u_{\mathrm{SDSS}} & = 0.483 x_S^3 - 1.341 x_S^2 + 1.168 x_S - 0.22, \\
		g_{\mathrm{KiDS}} - g_{\mathrm{SDSS}} & = -0.035 x_S + 0.038, \\
		g_{\mathrm{ATLAS}} - g_{\mathrm{SDSS}} & = -0.024 x_S + 0.015, \\
		g_{\mathrm{NSC}} - g_{\mathrm{SDSS}} & =  -0.067 x_S - 0.041, \\
		g_{\mathrm{KiDS}} - g_{\mathrm{Ref}} & = 0.073 x_R + 0.054, \\
		g_{\mathrm{ATLAS}} - g_{\mathrm{Ref}} & =  0.085 x_R + 0.026, \\
		g_{\mathrm{NSC}} - g_{\mathrm{Ref}} & =  0.044 x_R - 0.027, \\
		g_{\mathrm{KiDS}} - g_{\mathrm{Pan}} & = 0.071 x_P + 0.055, \\
		g_{\mathrm{ATLAS}} - g_{\mathrm{Pan}} & =  0.083 x_P + 0.027, \\
		g_{\mathrm{NSC}} - g_{\mathrm{Pan}} & =  0.042 x_P - 0.025, 
	\end{aligned}
	\label{eqn:color}
\end{equation}
where 
\begin{equation}
	\begin{aligned}
	x_S & = g_{\mathrm{SDSS}} -  i_{\mathrm{SDSS}}, \\
	x_R & = g_{\mathrm{Ref}} -  i_{\mathrm{Ref}}, \\
	x_P & = g_{\mathrm{Pan}} -  i_{\mathrm{Pan}}.
	\end{aligned}
\end{equation}
We capture only the bulk of the stellar population by rejecting approximately $5\%$ of stars on both the blue and red sides, leading to the color ranges listed in Table~\ref{tab:range}.
\begin{table}
	\caption{Color ranges of equations (\ref{eqn:color}) and (\ref{eqn:color_synth}). For (\ref{eqn:color}), The VST entries apply to both KiDS and ATLAS. The bounds are selected to capture the bulk of the population by rejecting $\sim 5$\% stars on both the blue and red sides from Fig.~\ref{fig:color}.}
	\label{tab:range}
	\centering
	\begin{tabular}{|r|c|}
		\hline
			                   &              color range  \\ \hline
		SDSS to KiDS           &  $ 0.3 < \left(g - i\right)_{\mathrm{SDSS}} < 1.5$ \\
		SDSS to ATLAS          &  $ 0.3 < \left(g - i\right)_{\mathrm{SDSS}} < 1.7$ \\
		SDSS to NSC            &  $ 0.3 < \left(g - i\right)_{\mathrm{SDSS}} < 1.2$ \\
		PanSTARRS to VST/NSC   &  $ 0.3 < \left(g - i\right)_{\mathrm{Pan}} < 1.5$ \\
		Refcat2 to VST/NSC     &  $ 0.3 < \left(g - i\right)_{\mathrm{Ref}} < 1.5$ \\
        \hline
    \end{tabular}
\end{table}

\begin{figure*}
    \includegraphics[width=\columnwidth]{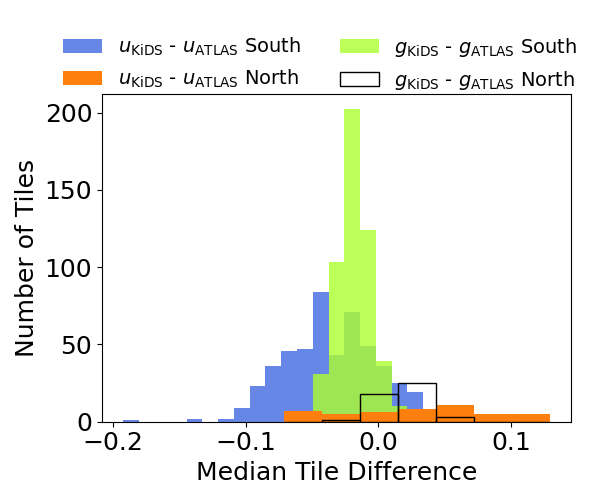}
    \includegraphics[width=\columnwidth]{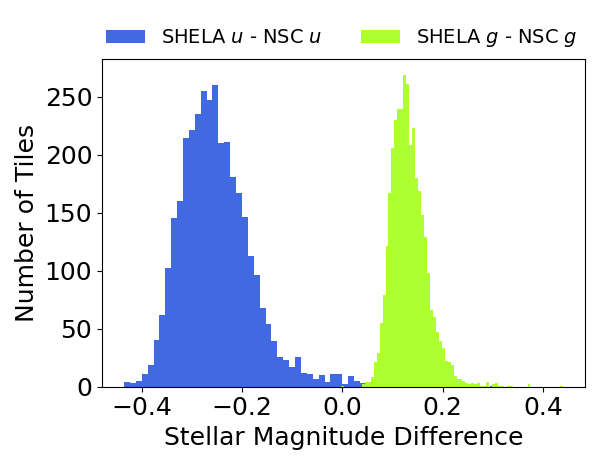}
    \includegraphics[width=1.23\columnwidth]{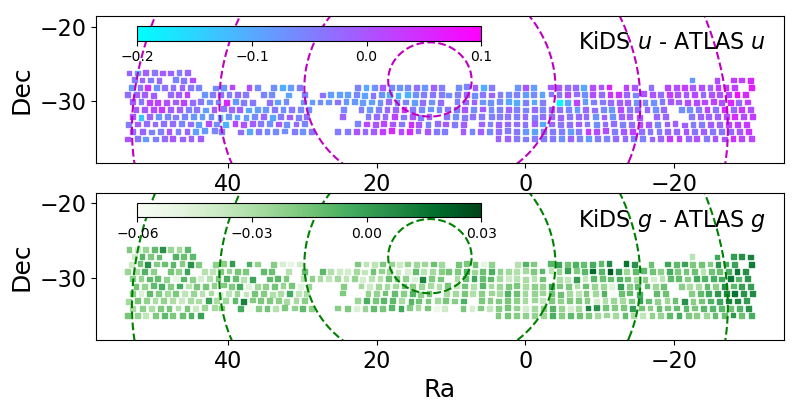}
    \includegraphics[width=0.8\columnwidth]{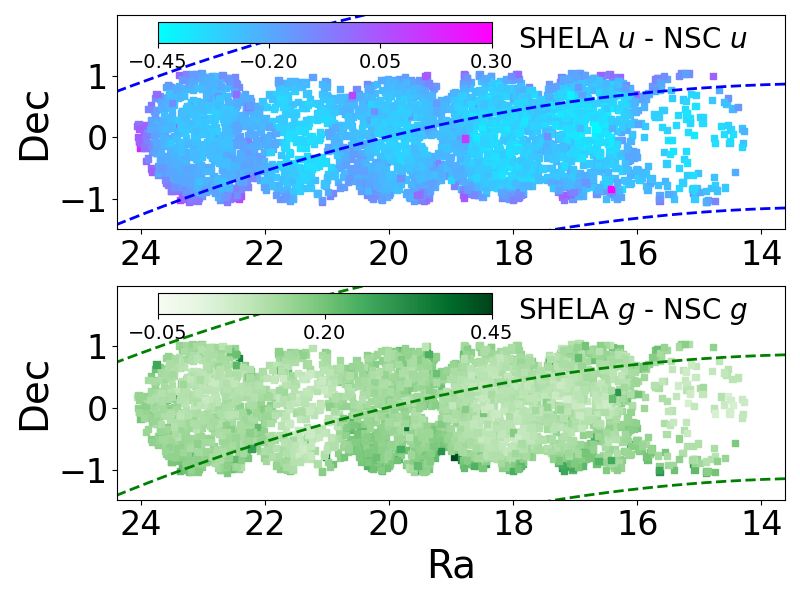}
	\caption{Upper Left: Median tile difference distribution between KiDS and ATLAS \textit{ug} bands. The $1\times 1$ deg$^2$ KiDS tiles  are matched with ATLAS. The median offset for the South tiles is $-0.035$ in the {\it u} band, and $-0.018$ in the {\it g} band. In the North, KiDS has only a few tiles that overlap ATLAS. They show similar median offsets, but in opposite sign to the South in both bands (KiDS fainter than ATLAS). Upper right: photometry difference between SHELA and NSC. Due to the low number of exposures, this panel shows the magnitude difference of stars. The overall median difference for all stars is $-0.26$ in the {\it u} band, and 0.13 in the {\it g} band. Lower: The spatial distribution of the histograms. The dashed lines indicate Galactic latitudes of [$-85^\circ, -75^\circ, -65^\circ, -55^\circ$] from inner-most to outside on the left panel, and [$-60^\circ, -62^\circ, -64^\circ$] from top to down on the right. Only the South tiles are shown for KiDS. We find no particular spatial pattern in all cases. Notice that the color-bars have different scales.}
	\label{fig:diff}
\end{figure*}

Notice that although KiDS and ATLAS are taken with the same telescope (VST) under the same set of filters, we find it necessary to fit the color equations separately. We return to this in App.~\ref{sec:appc}, but note that a direct comparison of measured KiDS/ATLAS {\it u}- and {\it g}-band magnitudes in the South (where the two surveys overlap) yields a {\it u}-band median offset of 0.035 mags and a {\it g}-band offset of 0.018 mags; the distribution is shown in Fig.~\ref{fig:diff}. The tile difference is calculated on the $1\times 1$ deg$^2$ KiDS tiles matched with ATLAS stars, with the cuts in previous sections applied. The large difference in some tiles in the \textit{u}-band is likely due to the different zero-point calibration adopted by the two surveys. Fig.~\ref{fig:diff} also shows the spatial distribution of offsets; which suggests a slight spatial variation with galactic latitude (recall that the KiDS {\it u}-band calibration also includes latitude-dependent correction).  Similarly, we find substantial median offsets between the SHELA \citep{shela} and the NSC photometry of the SHELA dataset, shown in the right panels of Fig.~\ref{fig:diff}. The median difference for all stars is found to be -0.26 in the \textit{u}-band and 0.13 in the \textit{g}-band. As noted before, SHELA is not part of our validation sets because the data are entirely included in the NSC catalog (albeit with a different photometric calibration to the SHELA data reduction).

These inconsistencies also show up in the constant terms in the color transform equations (\ref{eqn:color}), making them dependent on the original calibration of the validation set. Therefore, we opt to derive the color terms with synthetic magnitudes, as shown in App.~\ref{sec:appc}.

\section{Color Equations from Synthetic Magnitudes}
\label{sec:appc}

We use synthetic magnitudes to derive magnitude transform for all instruments involved in this work. We first assemble a library of stellar spectra that cover the \textit{ugi} band region:
\begin{enumerate}
    \item The X-Shooter Spectral Library \citep[XLS;][]{xsl_dr1, xsl_dr2}, containing 813 observations of 666 stars over the wavelength range $3000-25000$ {\AA} at a resolution R $\sim 10000$. The spectra were corrected for instrument transmission and telluric absorption, as well as wavelength-dependent flux losses. A comparison with other libraries (the MILES library, the combined IRTF and Extended IRTF) shows an accuracy to better than $<1$\% in the synthesized broad-band colors \citep{xsl_dr2}.
    \item The CALSPEC spectra library \citep{cal_rev, calspec}, consisting of the stellar spectra that are the spectrophotometric flux standards of the HST photometric system. All spectra were obtained with the Space Telescope Imaging Spectrograph (STIS) installed on HST that covers $1140-10200$ {\AA} (though some spectra are truncate in the \textit{u}-band region). Some sources also have infrared coverages from the NICMOS grism and the WFC3 grisms on HST.
    \item The Pickles spectra library \citep{pic}, a library of 131 flux calibrated spectra, covering all normal spectral types and luminosity classes, and a variety of metallicity abundance. The wavelength coverage is $1150-25000$ {\AA} with a resolution of $R=500$. The Pickles library contains sources where the atmospheric absorption features are not corrected, but simply set to zero. We note that this effect is minimal in the calculation of the synthetic magnitudes. Moreover, as the Pickles library only contribute a small amount to the total spectra used in this work, the effect is negligible.
\end{enumerate}

All of the 131 spectra from Pickles are used. The coordinate information of stars is missing in the Pickles library, so the extinction is not corrected. Since the Pickles stars are much brighter than the XSL/CALSPEC sources, and are thus much closer, the extinction causes only minor reddening that can be ignored.  Sources with low extinction E(B-V)<0.1 are selected from the XSL and the CALSPEC library. They are de-reddened according to the Fitzpatrick reddening laws following equation (A1) of \citet{spec_dered}. Since some of the XSL sources have large flux uncertainties, we propagate those uncertainties into the magnitude uncertainties, and select only those with $u_{\mathrm{err}}, g_{\mathrm{err}} < 0.002$. 606 XSL sources pass the selection. In addition, 91 CALSPEC sources with complete wavelength coverage in the \textit{ugi} bands are selected. This yields a total of 828 spectra. 

\begin{figure*}
    \centering
        \includegraphics[width=0.99\hsize]{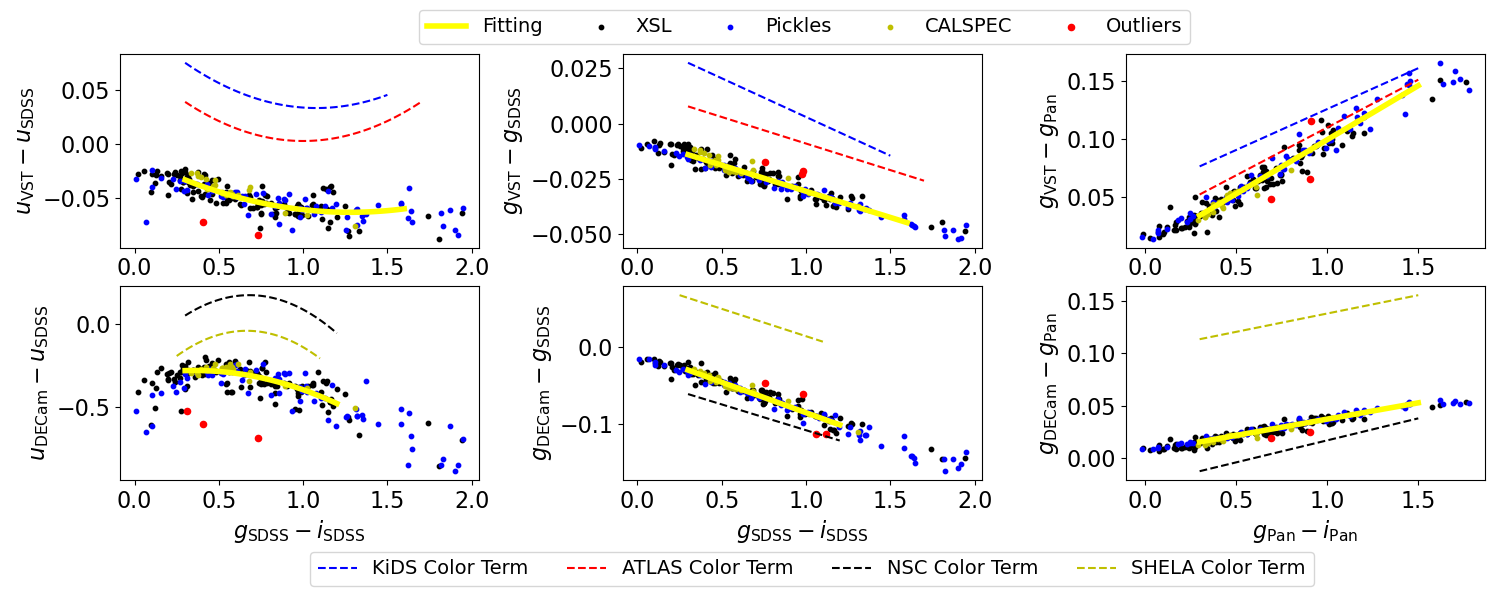}
	\caption{Magnitude transform fitted on synthetic magnitudes. A total of 828 spectra from Pickles, XSL and CALSPEC are used for the fitting. An iterative $3\sigma$ clipping is applied in the fitting process, with rejected source marked as red. The fitted color equations are shown in solid yellow curves, and can be found in equation (\ref{eqn:color_synth}). For comparison, the magnitude equations fitted on actual survey data (see App.~\ref{sec:appb} for details) are also plotted in dashed lines.}
    \label{fig:synth}
\end{figure*}

We obtain the filter transmission curves (including atmospheric transmission and system throughput) from 
the Filter Profile Service maintained by the Spanish Virtual Observatory\footnote{\url{http://svo2.cab.inta-csic.es/theory/fps/}},
and use these to compute synthetic broad-band magnitudes for all datasets in question, and fit for the
color equations (shown in Fig.~\ref{fig:synth}). The best-fit color equations are:
\begin{equation}
	\begin{aligned}
		u_{\mathrm{VST}} - u_{\mathrm{SDSS}} & = 0.031 x_S^2 - 0.08 x_S - 0.012, \\
		u_{\mathrm{DECam}} - u_{\mathrm{SDSS}} & = -0.28 x_S^2 + 0.197 x_S - 0.315, \\
		g_{\mathrm{VST}} - g_{\mathrm{SDSS}} & = -0.024 x_S - 0.007, \\
		g_{\mathrm{DECam}} - g_{\mathrm{SDSS}} & =  -0.079 x_S - 0.006, \\
		g_{\mathrm{VST}} - g_{\mathrm{Pan}} & = 0.094 x_P + 0.005, \\
		g_{\mathrm{DECam}} - g_{\mathrm{Pan}} & =  0.031 x_P + 0.006.
	\end{aligned}
	\label{eqn:color_synth}
\end{equation}
where 
\begin{equation}
    \qquad
	x_S = g_{\mathrm{SDSS}} -  i_{\mathrm{SDSS}} \quad \mathrm{and} \quad
	x_P = g_{\mathrm{Pan}} -  i_{\mathrm{Pan}}.
\end{equation}
The color ranges over which we fit (and thus the range these equations apply to) follows the bulk of photometry data in the validation sets, and are listed in Table~\ref{tab:range}. 

For comparison, the color equations derived from the actual survey data (equation \ref{fig:color}) are also shown in Fig.~\ref{fig:synth} in dashed lines. We find that all of them show (large) offsets to the synthetic magnitudes, but also that the color transforms determined from different survey data / calibration on the same instruments (i.e. KiDS vs. ATLAS on VST, and NSC. vs. SHELA on DECam) differ noticeably. This is likely the root cause of the inconsistencies we find in App.~\ref{sec:appb}.

For the above reasons, we choose to use color equations (\ref{eqn:color_synth}) for the re-calibration of the validation sets. This results in 1). an independent calibration regardless of the initial photometry of the validation sets, and 2). an accurate calibration that aligns with the SDSS/PanSTARRS AB system.
\section{Gaia parallax and proper motion}
\label{sec:appd}

We attribute the variation of the observed $u-g$ edge color in halo MSTO stars with galactic latitude to contamination from stars in the galactic disk. The Gaia satellite, currently in
Early Data Release 3 \citep[EDR3;][]{gaia_mission, gaia_edr3}, offers high precision parallax and proper motion measurements, providing a possible way of rejecting disk stars based on their Gaia distance / motion measurements. We here test the usage of Gaia information to reduce the contamination.

We retrieve stars from the Gaia Archive\footnote{\url{https://gea.esac.esa.int/archive/}} for RA, DEC, parallax and proper motion. For each star, the total proper motion uncertainty is calculated using
$$
\mathrm{pm\_err} = \frac{1}{\mathrm{pm}}\sqrt{{\tt pmra}^2 *{\tt pmra\_error}^2 + {\tt pmdec}^2* {\tt pmdec\_error}^2}
$$
where {\tt pmra}, {\tt pmra\_error} are the proper motion in right ascension direction  and its uncertainty, and {\tt pmdec, pmdec\_error} those in the declination direction.  We match these Gaia stars to the validation sets, and remove stars from the validation sets with a parallax cut according to the plane-parallel disk model:
$$
	\mathrm{parallax} - \mathrm{parallax\_error} > |\sin{b}|.
$$
This cut essentially removes stars within the disk at 1-sigma confidence (assuming that the disk is 1 kpc thick).
In addition, we place a proper motion cut, following the assumption that fast-moving stars are likely to be nearby and therefore in the disk:
$$
	\mathrm{pm} - \mathrm{pm\_err} > 10 \quad \mathrm{mas/year}.
$$
Stars that satisfy either threshold are rejected. We keep stars that don't have a Gaia counter-part or are lacking both the parallax and the proper motion measurements.

\begin{figure}
	\centering
	\includegraphics[width=\columnwidth]{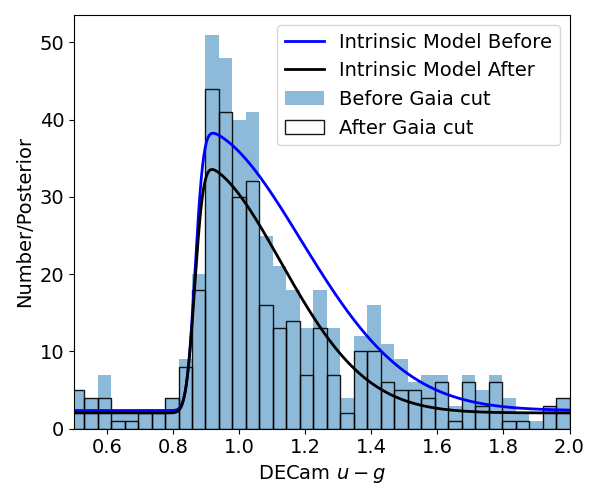}
    \caption{Stars rejected by Gaia from a $1.5\times1.5$ deg$^2$ DECam field. The blue histogram shows all the DECam stars within our selection (the ``white box'' selection of Fig.~\ref{fig:iso}). The empty histogram is the remaining stars after the parallax and proper motion cut of Gaia. The full model (\ref{eqn:conv}) is fitted on both histograms, and the resulting parameters are used for plotting the intrinsic model (\ref{eqn:intrinsic_model}) in this Figure No significant change in the blue tip color $x_0$ is observed after the Gaia cuts.}
	\label{fig:gaia}
\end{figure}

We examine three fields from different surveys in the validation sets at different galactic latitude, listed in Table~\ref{tab:gaia}. The blue tip color $x_0$ is measured before and after the Gaia cuts. We perform the same test for stars in the SDSS fields that overlap the three selected fields. In all cases, the Gaia cuts reject about 20 -- 25\% of stars in our selection, but barely alter the blue tip colors. A DECam field is shown in Fig.~\ref{fig:gaia} as an example. As a conclusion, Gaia EDR3 is not deep enough into the disk for rejecting the disk stars. We look forward to future Gaia data release for this purpose.

\begin{table}
	\caption{Testing the impact of Gaia cuts on the blue tip color. The blue tip colors of three fields at different galactic latitudes and from different surveys are listed, measured both without the Gaia cuts, and with the Gaia cuts. We repeat the test for the samples of SDSS stars overlapping these three fields. The last two rows list the shift in the measured edge color when applying the Gaia cuts - no significant change in the blue tip color is observed in all 6 cases.}
	\label{tab:gaia}
	\centering
	\begin{tabular}{l|c|c|c}
		\hline
		Survey             &       ATLAS      &       KiDS       &      DECam        \\ \hline
		Ra, Dec            &  60.14, -12.06   & 181.03, 1.49    &  187.49, 12.56     \\
        Galactic Latitude  &        -43.29    &        61.96     &        74.57         \\		
		\# White Box Stars &        214       &        236       &        473        \\
		\# Stars Gaia Rejected   &         37       &         51       &        115        \\
		$x_0$ w/o Gaia cuts      &  $0.79\pm 0.02$  &  $0.80\pm 0.02$  &  $0.87\pm 0.01$   \\
		$x_0$ w/ Gaia cuts       &  $0.79\pm 0.02$  &  $0.81\pm 0.02$  &  $0.87\pm 0.01$   \\
		$x_0$ SDSS w/o cuts  &  $0.76\pm 0.02$  &  $0.78\pm 0.01$  &  $0.76\pm 0.01$   \\
		$x_0$ SDSS w/ cuts   &  $0.75\pm 0.02$  &  $0.78\pm 0.01$  &  $0.76\pm 0.01$   \\
		$\Delta x_0$       &  $0.00\pm 0.03$  &  $0.01\pm 0.03$  &  $0.00\pm 0.01$   \\
		$\Delta x_0$ SDSS  &  $0.01\pm 0.03$  &  $0.00\pm 0.01$  &  $0.00\pm 0.01$   \\
		\hline
    \end{tabular}
\end{table}

\section{Comparison with Stellar Locus based Calibration}
\label{sec:appe}

\begin{figure}
	\centering
	\includegraphics[width=\columnwidth]{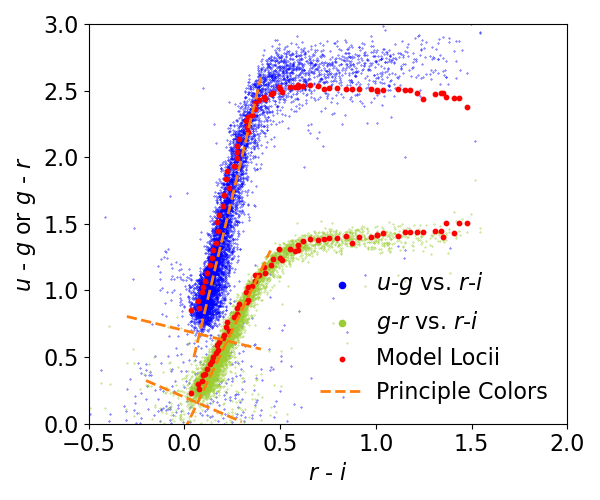}
    \caption{Illustration of the stellar locus, Big MACS and the \textit{s}-color method. The observed stars are plotted in blue and green for \textit{u-g} and \textit{g-r} respectively -- they largely form a one-dimensional sequence, the stellar locus. The red dots are a synthetic model locus taken from Big MACS. The dashed lines are axes of the ``principle colors'' defined to align with the linear part of the stellar locii.  The observed data in this plot is a $4\times4$ deg$^2$ KiDS tile. Note the large scatter in the \textit{u-g} color of the redder stars ($r-i > 0.5$) in the observed data, caused by metallicity variations in the sample stars.}
	\label{fig:slr}
\end{figure}

For completeness, we compare our calibration on SDSS data with two methods based on fitting the stellar locus: Big MACS by \cite{bigmacs,big-macs} and the \textit{s}-color method by \citet{s_color} and \citet{s_color_cali}. Both methods rely on the fact that stars have well-defined trajectories in the color-color spaces (the stellar locus, see Fig.~\ref{fig:slr}).
In the following, we apply the two methods to each of the 640 SDSS fields used in this work (Sect.~\ref{sec:data}), excluding a few outliers identified in Fig.~\ref{fig:pos}.  For each field, we pre-select stars in the same way as described in  Sect.~\ref{subsec:sdss_data}.

Big MACS (and its predecessor {\tt Stellar Locus Regression, SLR} by \citeauthor{slr_3} \citeyear{slr_3}) compares observed colors with empirical model stellar locii based on an SED library from SDSS and Pickles, and a set of system responses.  It simultaneously solves for the zeropoints of all bands relative to an ``anchor band''.  A precise calibration of colors is possible through the inclusion of the full stellar locii including the ``kinks'' (see Fig.~\ref{fig:slr}). In the calibration process with Big MACS, we keep the \textit{r}-band ZP fixed and fit for the \textit{ugi} ZPs.

\begin{figure}
	\includegraphics[width=0.9\columnwidth]{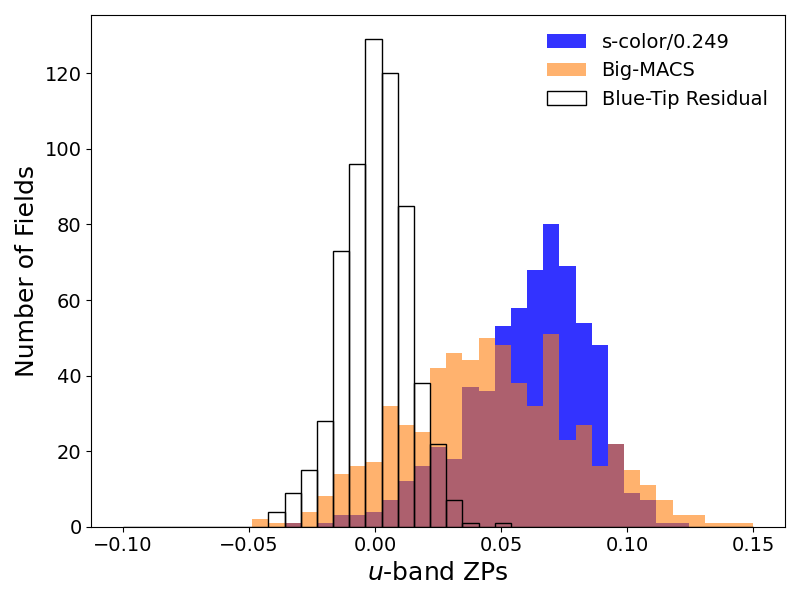}
	\caption{Zero-Point distribution of SDSS \textit{u}-band using stellar locus based methods. For the Big-MACS method, we fix \textit{r}-band as the anchor and calibrate \textit{ugi}-bands simultaneously. For the \textit{s}-color method, we assume the SDSS \textit{gr}-bands to be well calibrated, and use the \textit{s}-color as the \textit{u}-band ZP.  The model residual of the blue-tip method (equation [\ref{equ:fit}] and Fig.~\ref{fig:pos}) is shown in the empty histogram as a comparison.}
	\label{fig:slr_hist}
\end{figure}

\begin{figure*}
    \includegraphics[width=0.9\columnwidth]{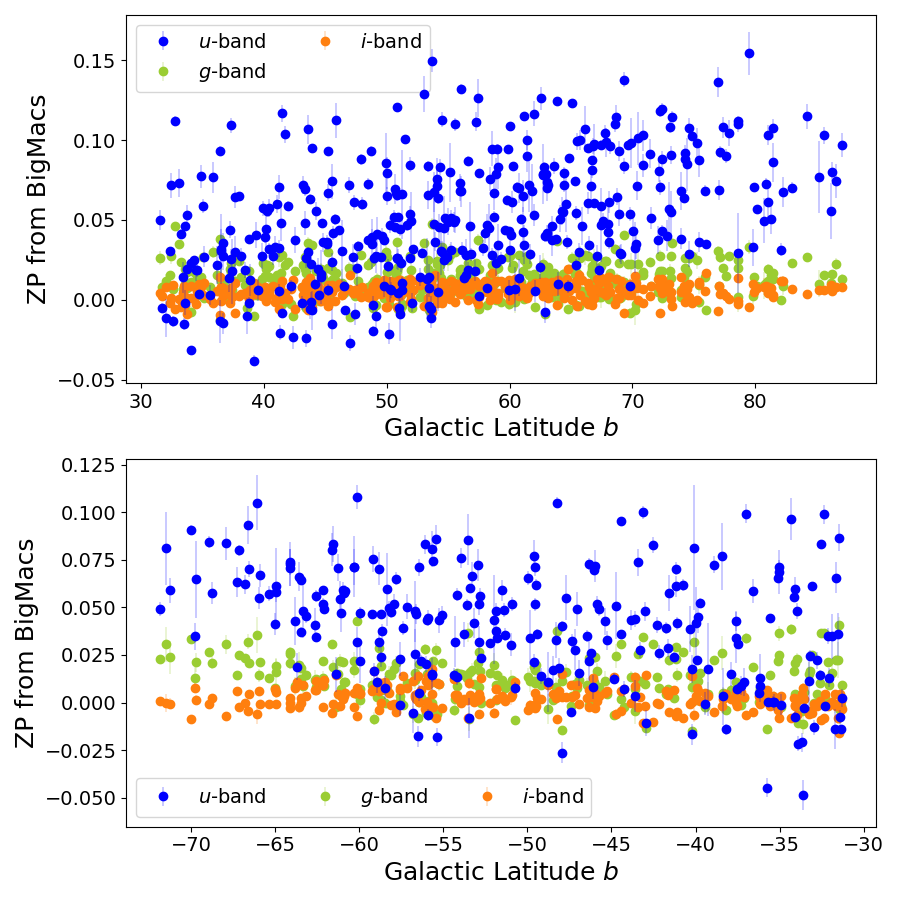}
	\includegraphics[width=0.9\columnwidth]{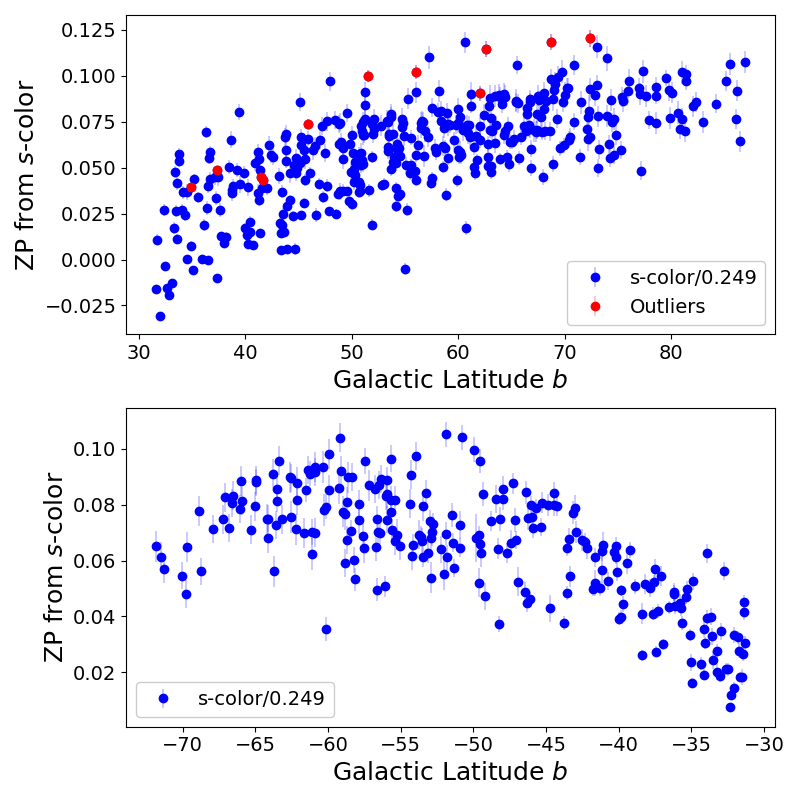}
	\caption{Location dependence of the \textit{u}-band ZPs. The left panel shows Big MACS ZPs for the same SDSS fields. The \textit{ugi}-bands ZPs are fitted while the \textit{r}-band is kept fixed as an anchor. Again we observe a latitude dependence on the \textit{u}-band ZPs. Also notice that the \textit{u}-band ZPs have a larger scatter than the other bands due to metallicity variation. The right panel shows the median \textit{s} color of each SDSS field (scaled down by 0.249 to be used as the \textit{u}-band ZPs) as a function of Galactic Latitude, where a clear trend can be observed. The red points are the same outliers as identified in Fig.~\ref{fig:pos}, and are excluded in both calibration methods. }
	\label{fig:slr_cali}
\end{figure*}

A histogram of the resulting Big MACS \textit{u}-band ZPs for the 640 SDSS fields is shown in orange in Fig.~\ref{fig:slr_hist}.  We find an RMS scatter of 0.034 mags around a mean value of 0.046 mags.  In comparison, we show the \textit{u}-band ZPs of the blue-tip method as empty histograms in Fig.~\ref{fig:slr_hist}; the RMS scatter is substantially smaller with 0.013 mags.  This is not surprising, as both \cite{slr_3} and \cite{bigmacs} warn that stellar locus regression is not applicable in the {\it u}-band due to the widening of the stellar locus due to metallicity variations (see Fig.~\ref{fig:slr}).  We compute the Big MACS zeropoints in {\it g-} and {\it i-}band and find much better performance with RMS values of 0.011 mags and 0.006 mags respectively.  In Fig.~\ref{fig:slr_cali} we show the Big MACS ZPs (of all \textit{ugi}-bands) as a function of galactic latitude - there is a clear trend for \textit{u}-band. We note that this trend is not to be directly compared with the trend in MSTO $u-g$ colors (Fig.~\ref{fig:pos}), as the Big MACS zeropoints are derived with all observed stars belonging to different populations: blue halo stars, blue disk stars and red disk stars. It is beyond the scope of this work to analyze their relative contribution to the zeropoints and model their relation to position. However, it is worth pointing out that for the photometric calibration of the J-PLUS survey, \citet{jplus_cali} combined model stellar locii with an empirical fitting on galactic location to achieve an 18 mmag calibration for \textit{u}-band.

The \textit{s}-color method can be loosely understood as a coordinate transform under which the (linear part of the) stellar locii are aligned with the coordinate axes (referred to as the ``principle colors'').  Unlike Big MACS and {\tt SLR}, the \textit{s}-color does not exclude photometry in the {\it u}-band.
We follow \citet{s_color_cali} to calculate the \textit{s}-color for SDSS stars in each field as
\begin{equation}
    \qquad s = -0.249 u_{_{\mathrm{SDSS}}} + 0.794 g_{_{\mathrm{SDSS}}} - 0.555 r_{_{\mathrm{SDSS}}} + 0.234,
    \label{eq:s-color}
\end{equation}
and select stars with
\begin{equation}
    \qquad \qquad \qquad r_{_{\mathrm{SDSS}}} < 19 \quad \mathrm{and} \quad -0.2 < s < 0.8.
\end{equation}
By construction, $s=0$ with perfect photometry, 
\begin{equation}
    \qquad 0 = -0.249 u_{_{\mathrm{true}}} + 0.794 g_{_{\mathrm{true}}} - 0.555 r_{_{\mathrm{true}}} + 0.234.
    \label{eq:s-color-true}
\end{equation}
We assume that the \textit{g} and \textit{r} bands are better calibrated while the \textit{u} band has larger zeropoint biases,
\begin{equation}
    \qquad \qquad u_{_{\mathrm{SDSS}}} + \delta u = u_{_{\mathrm{true}}}, \quad
    g_{_{\mathrm{SDSS}}} \approx g_{_{\mathrm{true}}}, \quad  r_{_{\mathrm{SDSS}}} \approx r_{_{\mathrm{true}}}.
\end{equation}
Then, the field-by-field scatter in the \textit{s}-color is directly linked to the biases in the \textit{u}-band zeropoints:
\begin{equation}
    \qquad s = -0.249 (u_{_{\mathrm{true}}} - \delta u) + 0.794 g_{_{\mathrm{true}}} - 0.555 r_{_{\mathrm{true}}} + 0.234,
\end{equation}
or
\begin{equation}
    \qquad \qquad \qquad \delta u = s/0.249.
    \label{eq:s-color-zp}
\end{equation}

The blue histogram in Fig.~\ref{fig:slr_hist} shows the distribution of thus determined median \textit{s}-color per field, re-scaled by a factor of -0.249 to recover the \textit{u}-band ZP precision. The width of the \textit{s}-color distribution (i.e. the RMS scatter around the mean value of 0.061) is 0.024 mags.  While smaller than the Big MACS value, the scatter is still substantially larger than the RMS model residual (0.013 mags) of the blue-tip method (empty histograms in Fig.~\ref{fig:slr_hist}). 

In Fig.~\ref{fig:slr_cali} we show the \textit{s}-color-derived zeropoints as a function of galactic latitude, and observe a clear trend again. We note that this trend is different from either the MSTO $u-g$ colors or the Big MACS zeropoints, as the \textit{s}-color (and thus the derived ZP) is a weighted combination of \textit{ugr}-bands instead of a simple measurement of stellar colors. This adds even more complexity to the interpretation of the trend.

As a conclusion, we find a $\sim2-3$\% calibration precision for the \textit{u}-band using stellar locus based methods, consistent with  \citet{big-macs} and \citet{s_color_cali}. In comparison, our disk contamination model (equation [\ref{equ:L}]) yields a model intrinsic scatter of 0.01 (Table.~\ref{tab:sdss}) as well as a residual total scatter of 0.013 mags (this section). It is plausible that these methods could benefit from modeling the Galactic latitude dependence similarly to what we have done here for the blue-tip method.

\bsp	
\label{lastpage}
\end{document}